\documentclass{aa}

\usepackage{graphicx}
\usepackage{longtable}

\begin{document}

\title{Host galaxy subtraction of TeV candidate BL Lacertae objects}

\author{K. Nilsson\inst{1}
\and M. Pasanen\inst{1}
\and L. O. Takalo\inst{1}
\and E. Lindfors\inst{1,2}
\and A. Berdyugin\inst{1}
\and S. Ciprini\inst{1,3}
\and J. Pforr\inst{4}
}

\offprints{K. Nilsson \email kani@utu.fi}

\institute{Tuorla Observatory, University of Turku, V\"ais\"al\"antie 20,
FIN-21500 Piikki\"o, Finland \and
Mets\"ahovi Radio Observatory, Helsinki University of Technology,
02540 Kylm\"al\"a, Finland \and
INFN Perugia \& Physics Dept., University of Perugia, via Pascoli, 
06123 Perugia, Italy \and
Landessternwarte Heidelberg, K\"onigstuhl, 69117 Heidelberg, Germany
}

\date{Received; accepted}

\abstract { 
Photometric monitoring of active galactic nuclei is often
complicated by the presence of a strong host galaxy component, which
adds unwanted flux to the measurement and introduces a
seeing-dependence to the flux that can plaque e.g. microvariability
studies.  We are currently monitoring a sample of 24 TeV candidate BL
Lacertae objects, many of which exhibit a prominent host galaxy
component, using differential aperture photometry.
} { 
In order to
study our light curves free from the above effects, we have derived the
host galaxy flux in differential aperture photometry as a function of
aperture radius and FWHM for 20 resolved sources in our sample.
}
{ 
We created accurate surface brightness
models of the targets and any significant nearby sources using
high-resolution R-band imaging obtained at the Nordic Optical
Telescope (NOT) and performed differential aperture photometry of the
models over a grid of aperture radii and FWHM values.
}  
{
The results are given as correction tables, that list the fluxes (in
mJy) of all ``contaminating'' sources (host galaxy + significant
nearby objects) as a function of aperture radius and FWHM. We found
that the derived fluxes depend strongly on aperture radius, but the
FWHM has only a minor effect (a few percent). We also discuss
the implications of our findings to optical monitoring programs and
potential sources of error in our derived fluxes. During this work we
have also constructed new calibration star sequences for 9 objects
and present the finding charts and calibrated magnitudes.
}
{}


\maketitle

\section{Introduction}

BL Lacertae objects (BL Lacs) are a class of active galactic nuclei
(AGN) characterized by weak or absent emission lines, variability of
flux over the whole electromagnetic spectrum and high optical and
radio polarization. Together with the flat spectrum radio quasars
(FSRQs) they form the group of blazars (e.g. Urry \& Padovani
\cite{urrpad95}). The spectral energy distribution of blazars has two
broad peaks, one at the mm--UV region and another at the x-ray--gamma
ray region (e.g. Fossati et al. \cite{fos98}). The lower energy peak
is very likely synchrotron radiation arising from relativistic
electrons spiralling in a magnetic field.  Various models have been
proposed for the origin of the high energy peak. The most popular
models involve inverse Compton (IC) scattering of soft photons from
the same electrons that produce the synchrotron radiation.
(e.g. Maraschi et al. \cite{mar92}, Bloom \& Marscher
\cite{bloo96}). These models predict correlations and time delays
between the low- and high-energy emission, thus emphasizing the
importance of long-term flux monitoring covering a wide range of
frequencies.

In the optical-NIR region the monitoring efforts are sometimes
hampered by the presence of a strong host galaxy component and/or
nearby companions.  Results obtained both from the ground (Falomo
\cite{fal96}; Heidt et al. \cite{heidt99}; Falomo \& Kotilainen
\cite{fk99}; Pursimo et al. \cite{pur02}; Nilsson et al. \cite{nil03})
and using the HST (Urry et al. \cite{urr98}; Urry et al. \cite{urr00};
Falomo et al. \cite{fal00}; Scarpa et al. \cite{sca00b}) have shown
the host galaxies of BL Lacertae objects to be bright (M$_{\rm R}$
$\sim$ -23.1\footnote{Throughout the paper we use $H_0 = 70$ km
s$^{-1}$ Mpc$^{-1}$, $\Omega_m$ = 0.3 and $\Omega_{\Lambda}$ = 0.7}),
large (r$_{\rm e}$ $\sim$ 7 kpc) and fairly round ($\epsilon$ =
0.1-0.3) elliptical galaxies, whose bulk properties do not differ
significantly from the population of inactive giant ellipticals.  The
luminosity evolution of the BL Lacertae host galaxies seems to be
consistent with passive evolution from a fairly distant (z = 2-3)
formation epoch (Heidt et al. \cite{hei04}; O'Dowd \& Urry
\cite{odowd05}), but these results need to be confirmed with more
complete samples.

Many BL Lacs also seem to have nearby companion galaxies at small ($<$
50 kpc) projected distances (e.g. Falomo \cite{fal96}; Falomo \&
Ulrich \cite{falul00}), but in most cases there is no spectroscopic
confirmation of real physical association between the BL Lac host and
its companion. Like in the case of quasars, it has been speculated
that tidal interaction between the BL Lac host galaxy and a companion
galaxy might have triggered the nuclear activity (e.g.  Heidt et
al. \cite{heidt99}). In some objects clear signs of tidal interaction
can be seen (e.g. 3C 371, Stickel et al. \cite{sti93}), but this
scenario needs be tested quantitatively with carefully selected
samples and proper control samples. One argument against the
interaction scenario is that BL Lac host galaxies appear very
symmetric with no obvious signs of recent strong interaction.

The optical fluxes of BL Lacs are most commonly measured from CCD
images in differential photometry mode, i.e. by comparing the BL Lac
brightness to the brightness of several calibrated comparison stars in
the field of the object.  In this way changes in sky transparency
can be eliminated and accuracies $<$ 1\% attained even under varying
conditions.  The images are usually measured using aperture
photometry, i.e. by integrating the light inside a circular aperture,
whose radius is determined by the angular extent of the source and by
the FWHM. Howell (\cite{how89}) determined the optimal extraction
aperture (i.e. the one that maximizes S/N) for point sources to be
$r_{opt}$ $\sim$ FWHM and to depend slightly on the brightness of
object on the CCD.

The presence of a strong host galaxy component can affect aperture
photometry at least in two ways.  Firstly, the host galaxy adds flux
to the measurement aperture. Since the amount of the additional flux
depends on the aperture radius, it is difficult to compare
observations made using different apertures.  In some cases the host
galaxy contribution can exceed the nuclear flux by a large margin.  In
these cases the broadband spectra are highly distorted and relative
flux variations underestimated in the optical region, unless a
correction for the host galaxy contribution is determined.

Secondly, as discussed by Carini et al. (\cite{car91}) and Cellone et
al. (\cite{cell00}), FWHM changes during the observations may
introduce false variability in objects with a prominent host galaxy
component. This is due to the fact that stars and galaxies have
different surface brightness profiles and thus they respond
differently to changes in the FWHM.  This effect is generally not very
strong (0.01-0.03 mag; Cellone et al. \cite{cell00} and this work),
but it may nevertheless affect studies that look for very low-level
intranight variability. The seeing effect can be further amplified by
nearby stars and galaxies, whose flux may leak into the measurement
aperture in a seeing-dependent manner.

Since the nearby environments and the relative contribution of the
host galaxy vary from one source to another, it is impossible to give
a general formula for the aperture photometry correction.  Instead,
each case has to be studied separately. The goal of this study is to
derive accurate aperture and seeing corrections for a sample of BL
Lacertae objects that we are currently monitoring at Tuorla
Observatory, taking also into account possible nearby stars and/or
companion galaxies. We also want to derive some general conclusions
with respect to optimizing aperture photometry for host galaxy
contaminated objects. In section 2 we will describe our sample and in
section 3 our analysis method in more detail.  Section 4 gives the
results of our analysis and section 5 summarizes our findings.

\section{The sample}

We are currently running an R-band monitoring program at Tuorla
Observatory of a sample of 24 BL Lacertae objects, comprising of all
sources in Costamante \& Ghisellini (\cite{cosghi02}) with $\delta >
+20^{\circ}$. These sources were selected as likely TeV emitters by
Costamante \& Ghisellini (\cite{cosghi02}), based mainly on their
location in the radio -- X-ray flux plane. With the advent of new
large collection area \v{C}erenkov telescopes (e.g. MAGIC;
Lorenz \cite{lor04} and HESS; Hofmann et al. \cite{hof01}) it is likely
that these sources will be targets of extensive monitoring
campaigns. Thus we started in fall 2002 to monitor these sources in
the R-band with the aim of providing a long-term light curve to better
constrain their variability characteristics and to alert the
MAGIC-telescope of a possible high state of the sources (see Albert et
al. \cite{alb06} for a successful trigger).  Many of the sources
in the list have been extensively monitored in the past decades
(e.g. BL Lac, Mrk 501 and OJ 287), whereas for some of the sources
only scattered points exist in the literature (e.g. 1ES 0033+595 and
RGB J0136+391).

The monitoring is made by obtaining CCD-images of the sources using
Tuorla 1.03 m and KVA 0.35 m telescopes and measuring the source and
suitable comparison stars with aperture photometry. The aperture
radius is fixed for each source to 4.0 - 7.5 arcsec depending
on the brightness of the source.  The flux of the source is obtained
via normal differential photometry using one of the comparison stars
as a primary reference and the rest as a check of the stability of the
comparison star and general quality of the photometry. Calibrated
R-band magnitudes are obtained by using published comparison star
magnitudes or comparison sequences calibrated by ourselves (see
Appendix A). The magnitudes are also corrected for the color
difference between the object and the comparison star using the
calibrated color coefficients of the telescopes.

\begin{table}
\caption{\label{newobs}An observing log of the new observations.}
\centering
\begin{tabular}{llll}
\hline\hline
Object & Date & t$_{\rm exp}$ & FWHM\\
       &      & [s]         & [arcsec]\\
\hline
1ES~1028+511 & 09 Feb 2007 & 400 & 0.70\\
RGB~J1117+202 & 08 Feb 2007 & 600 & 0.72\\
1ES~1426+428 & 08 Dec 2007 & 720 & 0.64\\
1ES~1544+820 & 27 Sep 2005 & 720 & 1.8\\
OT~546       & 27 Sep 2005 & 720 & 2.1\\ 
\hline
\end{tabular}
\end{table}

\begin{table*}[t]
\caption{ 
The photometric parameters adopted for the host galaxies and nearby
companions of the objects in this study. Column 3 gives the FWHM of
the observation and columns 5 and 6 give the offset (in arcsec)
relative to the BL Lacertae nucleus. Values listed in parentheses
were held constant during the fitting (see Section 3 for details
of the modeling).
}
\label{hostidata}
\begin{center}
\tiny
\begin{tabular}{lcclcccccccl}
\hline\hline
Object        & z & FWHM     & Type & $\Delta$x & $\Delta$y & R mag & r$_{\rm eff}$ & ell &  PA   &  $\beta$ &  ref.\\
              &   & (arcsec) &      & (arcsec)  & (arcsec)  &       & (arcsec)      &     & (deg) &          &\\   
\hline
1ES 0033+595  &       & 1.13 & star   &   1.4 & -0.7 & 17.88$\pm$0.03 &               &                &            &               &1,4$^1$\\
1ES 0120+340  & 0.272 & 1.04 & host   &   0.0 &  0.0 & 17.89$\pm$0.07 &   2.0$\pm$0.2 &  0.08$\pm$0.02 & 137$\pm$46 & 0.12$\pm$0.02 &4\\
              &       &      & star   &  -0.4 &  6.4 & 17.11$\pm$0.04 &               &                &            &               &1\\
RGB J0214+517 & 0.049 & 0.76 & host   &   0.0 &  0.0 & 14.08$\pm$0.07 &  17.4$\pm$1.1 &  0.15$\pm$0.01 & 132$\pm$1  & 0.12$\pm$0.01 &4\\
              &       &      & star   &  -6.1 &  0.3 & 17.14$\pm$0.06 &               &                &            &               &1\\
1ES 0806+524  & 0.138 & 0.99 & host   &   0.0 &  0.0 & 16.10$\pm$0.07 &   4.6$\pm$0.2 &  0.10$\pm$0.02 & 106$\pm$4  & 0.14$\pm$0.02 &1\\
1ES 1011+496  & 0.212 & 0.96 & host   &   0.0 &  0.0 & 16.41$\pm$0.09 &   4.0$\pm$0.6 &  0.28$\pm$0.03 &  42$\pm$4  & 0.06$\pm$0.01 &4\\
1ES 1028+511  & 0.360 & 0.70 & host   &   0.0 &  0.0 & 18.60$\pm$0.22 &   2.0$\pm$4.7 &  0.34$\pm$0.19 & 152$\pm$44 & (0.25)        &1\\
Mrk 421       & 0.031 & 0.53 & host   &   0.0 &  0.0 & 13.18$\pm$0.05 &   8.2$\pm$0.2 &  0.21$\pm$0.01 & 105$\pm$1  & 0.36$\pm$0.02 &3\\
              &       &      & galaxy$^2$ & -11.4 &  8.3 & 15.54$\pm$0.04 &   7.3$\pm$0.1 &  0.39$\pm$0.01 &  81$\pm$1  & (0.25)        &1\\
              &       &      & galaxy$^2$ & -11.4 &  8.3 & 16.51$\pm$0.04 &   3.0$\pm$0.1 &  0.79$\pm$0.01 &  66$\pm$1  & (1.00)        &1\\
RGB J1117+202 & 0.139 & 0.72 & host   &   0.0 &  0.0 & 16.31$\pm$0.13 &   4.6$\pm$0.6 &  0.22$\pm$0.03 & 162$\pm$5  & 0.28$\pm$0.07 &1\\
              &       &      & galaxy &  -8.1 & -3.7 & 18.23$\pm$0.05 &   0.9$\pm$0.1 &  0.11$\pm$0.03 &  94$\pm$8  & (0.25)        &1\\
              &       &      & galaxy &  10.8 &  4.1 & 17.79$\pm$0.06 &   2.3$\pm$0.1 &  0.33$\pm$0.02 &  75$\pm$2  & 0.38$\pm$0.02 &1\\
              &       &      & star   &   4.8 & 10.0 & 19.99$\pm$0.05 &               &                &            &               &1\\  
Mrk 180       & 0.045 & 1.17 & host   &   0.0 &  0.0 & 14.08$\pm$0.05 &   8.3$\pm$0.3 &  0.05$\pm$0.01 &  15$\pm$3  & 0.19$\pm$0.01 &4\\
              &       &      & star   &  -0.6 & -6.3 & 15.21$\pm$0.04 &               &                &            &               &1\\
RGB J1136+676 & 0.135 & 0.96 & host   &   0.0 &  0.0 & 15.93$\pm$0.05 &   4.1$\pm$0.2 &  0.13$\pm$0.01 & 130$\pm$3  & 0.16$\pm$0.01 &4\\
              &       &      & galaxy &   6.4 &  6.0 & 19.14$\pm$0.32 &   0.6$\pm$0.3 &  0.09$\pm$0.08 & 162$\pm$54 & 0.13$\pm$0.21 &1\\
ON 325        & 0.130 & 1.10 & host   &   0.0 &  0.0 & 15.26$\pm$0.32 &  19.4$\pm$6.2 &  0.11$\pm$0.05 &  13$\pm$22 & 0.04$\pm$0.06 &4\\
              &       &      & galaxy &   2.1 & -2.5 & 18.90$\pm$0.05 &   0.5$\pm$0.1 &  0.40$\pm$0.03 &  89$\pm$2  & (0.25)        &1\\
1ES 1218+304  & 0.182 & 1.61 & host   &   0.0 &  0.0 & 16.86$\pm$0.03 &   3.6$\pm$0.3 &  0.24$\pm$0.02 &  88$\pm$2  & (0.25)        &1\\
              &       &      & star   & -15.1 &  1.3 & 15.59$\pm$0.03 &               &                &            &               &1\\
RGB J1417+257 & 0.237 & 0.54 & host   &   0.0 &  0.0 & 16.66$\pm$0.06 &   5.2$\pm$0.3 &  0.14$\pm$0.02 & 166$\pm$3  & (0.25)        &1\\
              &       &      & galaxy &  1.8  & -4.8 & 18.14$\pm$0.23 &   2.6$\pm$0.7 &  0.34$\pm$0.05 & 139$\pm$5  & 0.22$\pm$0.11 &1\\
1ES 1426+428  & 0.129 & 0.64 & host   &   0.0 &  0.0 & 15.93$\pm$0.05 &   4.2$\pm$0.2 &  0.37$\pm$0.01 & 118$\pm$1  & 0.21$\pm$0.02 &1\\
1ES 1544+820  &       & 1.83 & star   &  -0.1 & -4.1 & 17.88$\pm$0.03 &               &                &            &               &1\\
Mrk 501       & 0.034 & 0.72 & host   &   0.0 &  0.0 & 11.92$\pm$0.06 &  48.2$\pm$4.2 &  0.24$\pm$0.01 & 169$\pm$1  & 0.09$\pm$0.01 &3\\
OT 546        & 0.055 & 2.08 & host   &   0.0 &  0.0 & 15.48$\pm$0.08 &   4.9$\pm$0.4 &  0.14$\pm$0.03 &  40$\pm$7  & 0.24$\pm$0.03 &1\\
1ES 1959+650  & 0.047 & 0.67 & host   &   0.0 &  0.0 & 15.08$\pm$0.03 &   6.4$\pm$0.1 &  0.21$\pm$0.01 &  97$\pm$1  & 0.44$\pm$0.02 &1,2\\
              &       &      & star   &   6.5 &  8.2 & 16.94$\pm$0.03 &               &                &            &               &1\\
BL Lac        & 0.069 & 0.77 & host   &   0.0 &  0.0 & 15.05$\pm$0.08 &  10.4$\pm$0.9 &  0.43$\pm$0.01 &  44$\pm$1  & 0.19$\pm$0.02 &4\\
1ES 2344+514  & 0.044 & 0.63 & host   &   0.0 &  0.0 & 13.90$\pm$0.06 &  10.9$\pm$0.6 &  0.25$\pm$0.01 & 104$\pm$1  & 0.19$\pm$0.01 &3\\
              &       &      & star   &   9.0 & 10.3 & 15.24$\pm$0.05 &               &                &            &               &1\\
\hline
\end{tabular}
\end{center}     
$^1$References: (1) this work,
(2) Heidt et al. (\cite{heidt99}), (3) Nilsson et al. (\cite{nil99}), 
(4) Nilsson et al. (\cite{nil03}). \hfill\\
$^2$The bulge and disk component of the companion galaxy.
\end{table*}

\section{Analysis}

Of the various factors affecting aperture photometry of host galaxy
dominated sources, we will concentrate on the two most important
factors: the aperture radius and FWHM. There are other factors that
may contribute noise, such as aperture centering and sky determination
errors, but in the following we will assume that these factors are
controlled by e.g. using an appropriate symmetry clean algorithm in
the aperture centering and careful selection of the sky region.

There are at least two methods one can use to estimate the host galaxy
contribution. Firstly, one can fit two-dimensional models consisting
of an unresolved nuclear component and a host galaxy component to each
monitoring frame and use the fitted nuclear magnitude as the result.
This is rarely feasible, however, owing to too poor resolution and
too low signal to noise in typical monitoring frames, which makes it
impossible to properly characterize the host galaxy. In addition, the
noise inherent to the fitting process can increase the total noise
into unacceptable levels, especially for sources with strong host
galaxy components. Furthermore, nearby stars and galaxies may be
difficult to mask out from the fit. The advantage of this method is
that if the PSF is determined from the observed frame, the effect of
the PSF shape is canceled, unless it varies considerably over the
field of view.

In the second method one fits two-dimensional models to
high-resolution, high-S/N images of the sources and computes aperture
correction tables through simulated frames.  This method, in addition
to being computationally more effective, has the advantage that the
fitting process is far more accurate than in the case of fitting
single monitoring frames. Further advantage of this approach is that
through simulated frames one can gain insight on the relative
importance of the factors (aperture size, PSF shape) under
investigation. There is one drawback of this method with respect to
the first one: in real monitoring data there is a high variation of
PSF shapes due to e.g. optical imperfections and tracking
errors. Since it is impossible in practice to compute correction
tables for every possible PSF shape, one has to focus on the most
important features of the PSF.  This somewhat limits the accuracy of
the second method, but given its advantages, we have opted to use this
method instead the first one. Furthermore, as we will show below,
this method can produce corrections that are sufficiently accurate in
most practical situations.

Our method of estimating host galaxy contribution thus consists of
four steps: 1) obtain a deep high-resolution image of the object and
fit a two-dimensional nucleus + host galaxy model to it, 2) Create a
simulated image of the object field with a proper comparison star
sequence, nearby sources and the object {\em without} the nuclear
component, 3) select the PSF shape and FWHM and and convolve the image
with the selected PSF, 4) Perform aperture photometry of the simulated
image. Steps 3 and 4 are repeated over a range of apertures and
seeing conditions to build a correction table. Each of these steps is
described in more detail below.

\subsection{Observations}

We have collected high-resolution R-band images of all 24 sources in
our monitoring sample using the Nordic Optical Telescope (NOT) in
1995-2007.  These images form a very homogeneous set of images with
respect to instrumentation, resolution and S/N and they are thus ideal
for our purpose. For 16 objects the photometric parameters of the host
galaxies have been published elsewhere by us (see the references in
Table \ref{hostidata}).  Renato Falomo kindly provided us with images of 3
objects (1ES~0806+524, 1ES~1218+304 and RGB~J1417+257) published in
Falomo \& Kotilainen (\cite{fk99}) and we have obtained new images of
5 objects using the ALFOSC instrument at the NOT (see Table
\ref{newobs}).

\subsection{Host galaxy decomposition}

For maximum accuracy it is important to analyze all images in a
homogeneous manner. The 16 previously published images were analyzed
by us using homogeneous procedures and thus satisfy this requirement,
but the 8 new objects need to be reanalyzed before proceeding to the
next phase. Below we will give a short description of our host galaxy
analysis (see Nilsson et al. \cite{nil99} for details).

The two-dimensional model of the source is assumed to consist of two
components, an unresolved nucleus parameterized by its position
($x_c$,$y_c$) and magnitude $m_c$ and the host galaxy parameterized by
position ($x_g$,$y_g$), total magnitude $m_g$, major axis effective
(half-light) radius $r_e$, ellipticity $\epsilon$, position angle $PA$
and slope of the brightness profile $\beta$. The host galaxy surface
brightness $I(r)$ is assumed to follow the S\'{e}rsic law
\begin{equation}
I(r) = I(r_e){\rm dex} \left\{ -b_{\beta} \left[ \left(
\frac{r}{r_e} \right) ^{\beta} - 1 \right] \right\}\ ,
\end{equation}
where $b_{\beta}$ is a $\beta$-dependent constant. The model is
convolved with a PSF obtained from sufficiently bright stars in the
field. Prior to fitting any overlapping stars/galaxies are masked out
from the fit.  The 10 free parameters (3 for the core and 7 for the
galaxy) are then fitted to the observed image by minimizing the chi
squared between the observed image and the model using an iterative
Levenberg-Marquardt loop.

The same fitting procedure is also applied to any significant nearby
stars and/or galaxies, except that the stars consist of only the
nuclear component and the galaxies of only the galaxy component.  To
be considered significant a galaxy or a star must be closer than
20\arcsec to the source and brighter than $m_g + 4.0$.

\begin{figure*}
\centering
\includegraphics[width=4.5cm]{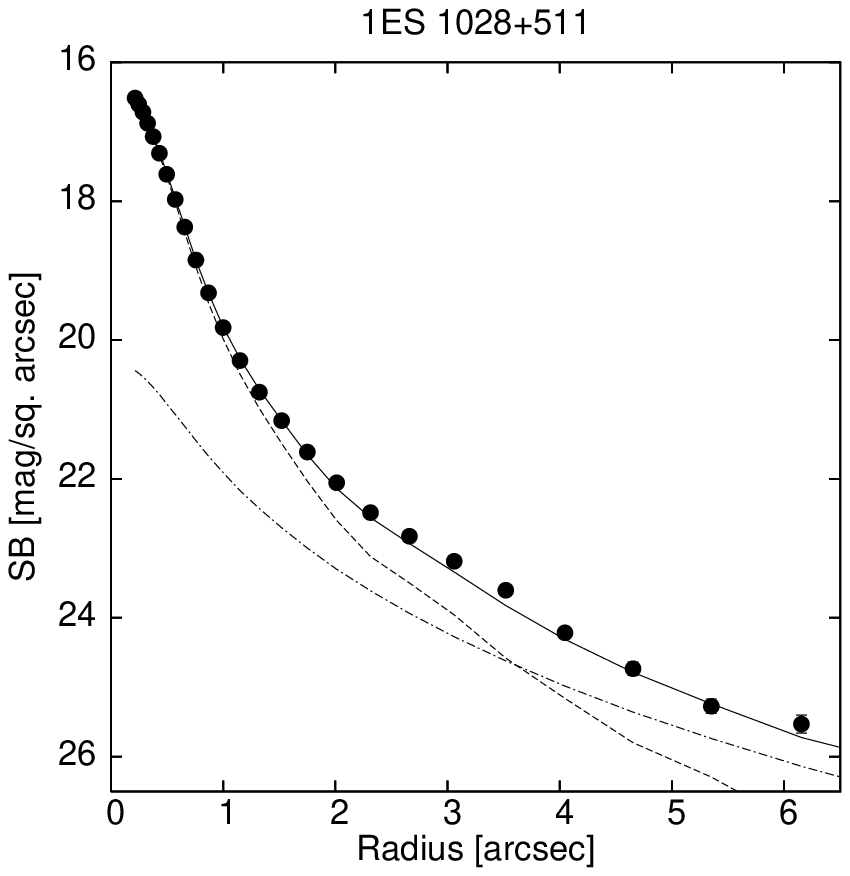}
\hspace*{10mm}
\includegraphics[width=4.5cm]{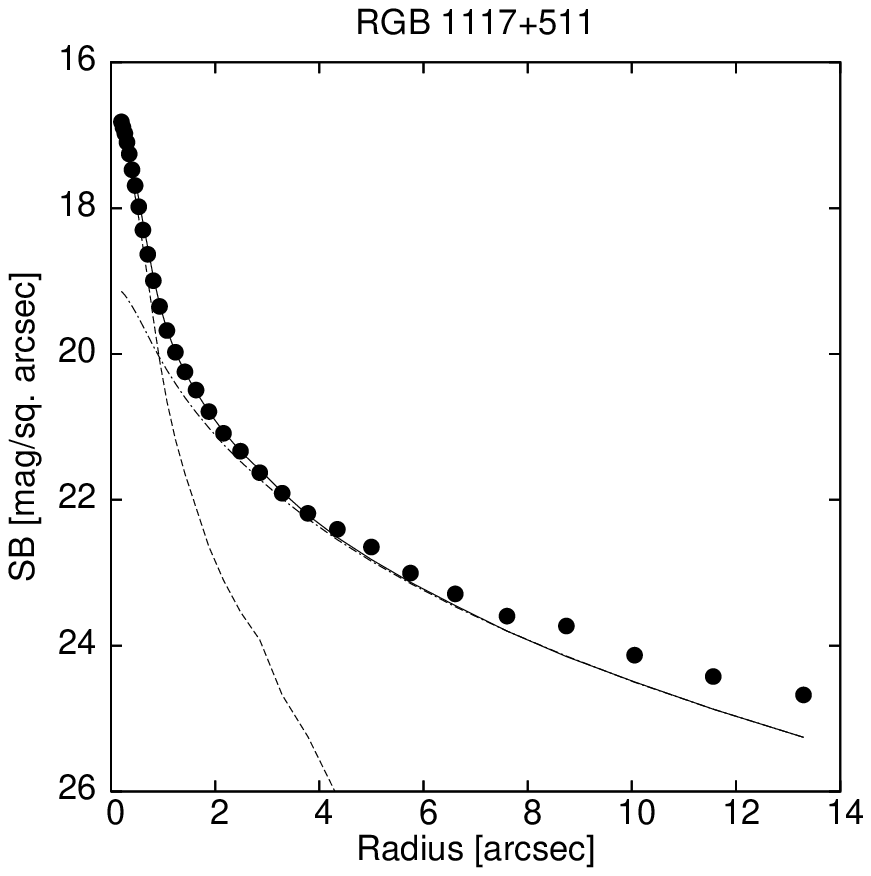}
\hspace*{10mm}
\includegraphics[width=4.5cm]{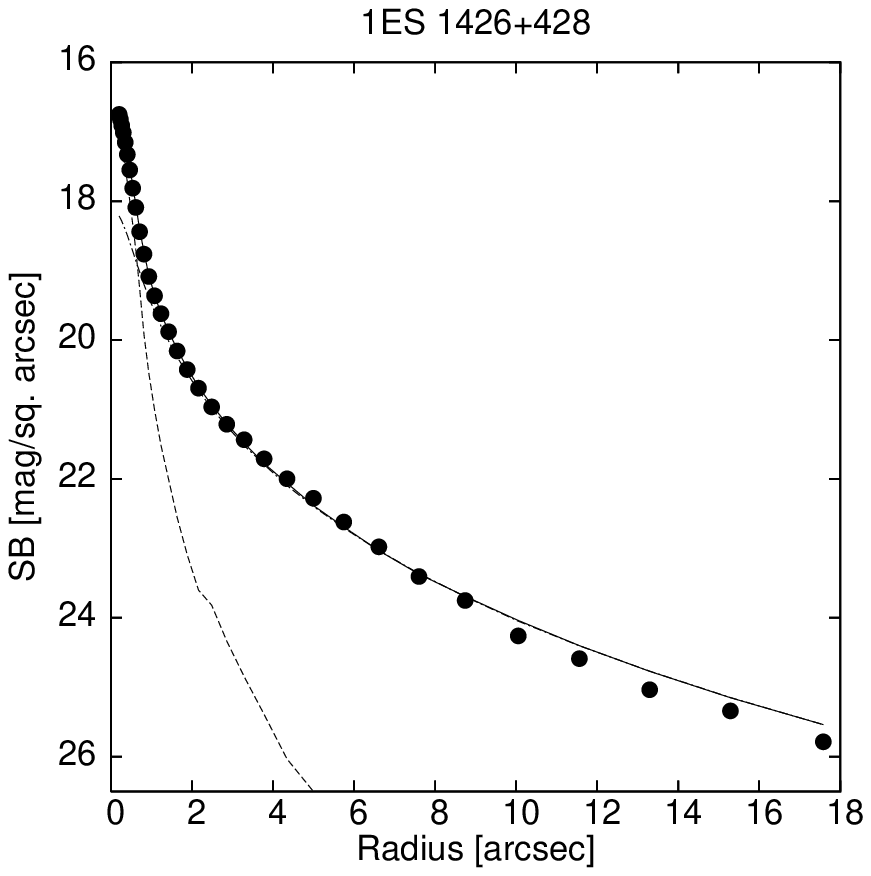}\\
\includegraphics[width=4.5cm]{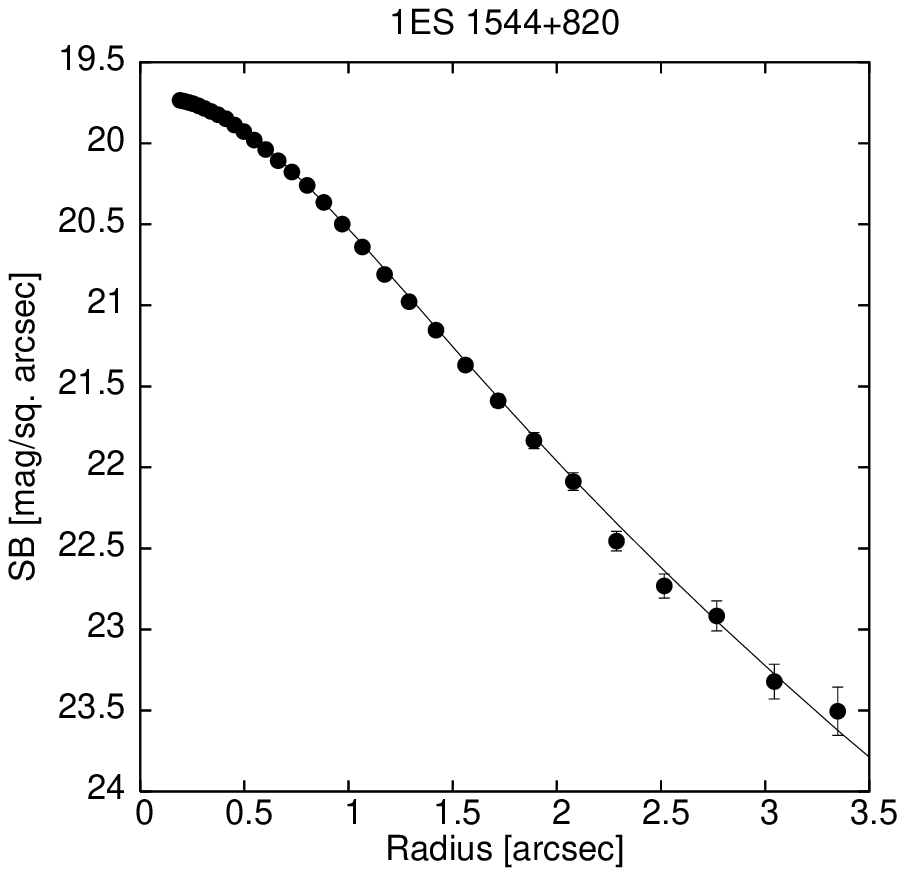}
\hspace*{11mm}
\includegraphics[width=4.5cm]{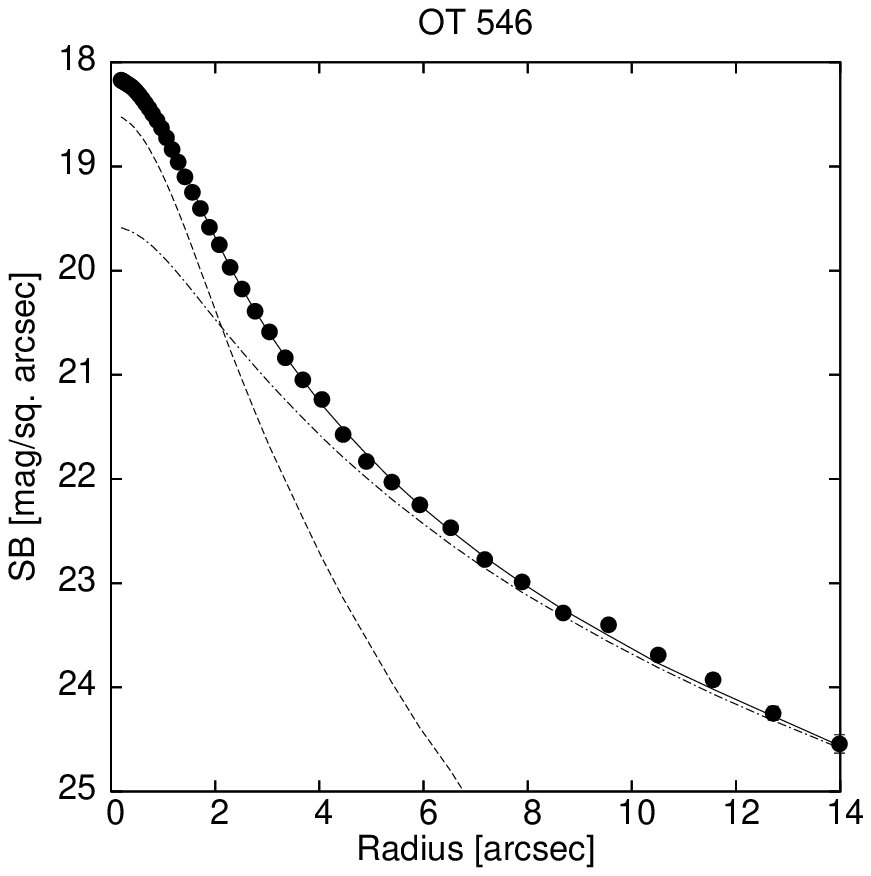}
\hspace*{5.8cm}
\caption{
\label{profiilit}
Surface brightness profiles of the newly observed sources. The
circles denote the observed surface brightness and the solid line the
modeled (core+host galaxy) surface brightness. The core and host galaxy
surface brightness are indicated by dashed and dot-dashed lines, 
respectively. Note that 1ES 1544+820 is unresolved and only a core-model
is shown.}
\end{figure*}

The results of the fitting procedure are given in Table
\ref{hostidata}.  We also show the surface brightness decomposition of
the newly observed sources in Fig. \ref{profiilit}.  As already noted
in Nilsson et al. (\cite{nil03}), the host galaxies are large ($r_e
\sim$ 7 kpc) and luminous (M$_{\rm R} \sim$ -23) elliptical galaxies,
whose surface brightness closely follows the S\'{e}rsic law.

Altogether 4 sources (1ES~0033+595, RGB~J0136+391, 1ES~0647+250 and
1ES~1544+820) were found to be completely unresolved with no signs of
the host galaxy, 2 (3C 66A and OJ 287) marginally resolved and 18
clearly resolved. In the two marginally resolved objects a faint host
galaxy component can be detected, but the significance of this
detection does not exceed the formal limit set in Nilsson et al.
(\cite{nil03}), namely that $r_e > 5 \sigma_{r_e}$, where
$\sigma_{r_e}$ is determined from simulations.  The host galaxy in
these two objects can have a small effect on aperture photometry, but
given the uncertainties in their host galaxy parameters we have left
them out in the following.  Note, however, that two of the unresolved
sources (1ES 0033+595 and 1ES 1544+820) have significant nearby stars
that need to be taken into account.  Our final list in Table
\ref{hostidata} thus incorporates 20 objects of the monitoring sample
with a significant host galaxy component and/or nearby companions.
Note that the host galaxy magnitudes in Table \ref{hostidata} slightly
differ from already published values due to the different calibration
adopted in this study, explained in more detail below.

\begin{table}
\centering
\caption{Adopted primary comparison star magnitudes.}
\label{primcompdata}
\begin{tabular}{lllcc}
\hline\hline
Object       & Star & R &  (V-R) & ref.$^1$\\
\hline
1ES 0033+595  &  D   & 13.66 $\pm$ 0.03 & 1.46 &1\\
1ES 0120+340  &  C   & 13.12 $\pm$ 0.03 & 0.38 &1\\
RGB J0214+517 &  A   & 13.85 $\pm$ 0.05 & 0.51 &1\\
1ES 0806+524  &  C2  & 14.22 $\pm$ 0.04 & 0.39 &3\\
1ES 1011+496  &  E   & 14.04 $\pm$ 0.03 & 0.39 &1\\
1ES 1028+511  &  1   & 12.93 $\pm$ 0.03 & 0.27 &5\\
Mrk 421       &  1   & 14.04 $\pm$ 0.02 & 0.32 &5\\
RGB J1117+202 &  E   & 13.56 $\pm$ 0.04 & 0.42 &1\\
Mrk 180       &  1   & 13.73 $\pm$ 0.02 & 0.25 &5\\
RGB J1136+676 &  D   & 14.58 $\pm$ 0.04 & 0.46 &1\\
ON 325        &  B   & 14.59 $\pm$ 0.04 & 0.37 &2\\
1ES 1218+304  &  B   & 13.61 $\pm$ 0.01 & 0.40 &4\\
RGB J1417+257 &  A   & 13.78 $\pm$ 0.04 & 0.57 &3\\
1ES 1426+428  &  B   & 14.17 $\pm$ 0.02 & 0.44 &4\\
1ES 1544+820  &  D   & 12.87 $\pm$ 0.03 & 0.31 &1\\
Mrk 501       &  4   & 14.96 $\pm$ 0.02 & 0.34 &5\\
              &  5   & 15.08 $\pm$ 0.02 & 0.43 &5\\
              &  6   & 14.99 $\pm$ 0.04 & 0.68 &5\\
OT 546        &  B   & 12.81 $\pm$ 0.06 & 0.33 &2\\
1ES 1959+650  &  4   & 14.08 $\pm$ 0.03 & 0.45 &5\\
              &  6   & 14.78 $\pm$ 0.03 & 0.42 &5\\
BL Lac        &  C   & 13.79 $\pm$ 0.05 & 0.47 &2\\
1ES 2344+514  &  C1  & 12.25 $\pm$ 0.04 & 0.36 &3\\
\hline
\end{tabular} 

$^1$ References: (1) this work, (2) Fiorucci \& Tosti (\cite{fio96}),
(3) Fiorucci et al. (\cite{fio98}), (4) Smith et al. (\cite{smi91}),
(5) Villata et al. (\cite{vill98}).
\end{table} 

\begin{figure*}[t]
\centering
\includegraphics[width=17cm]{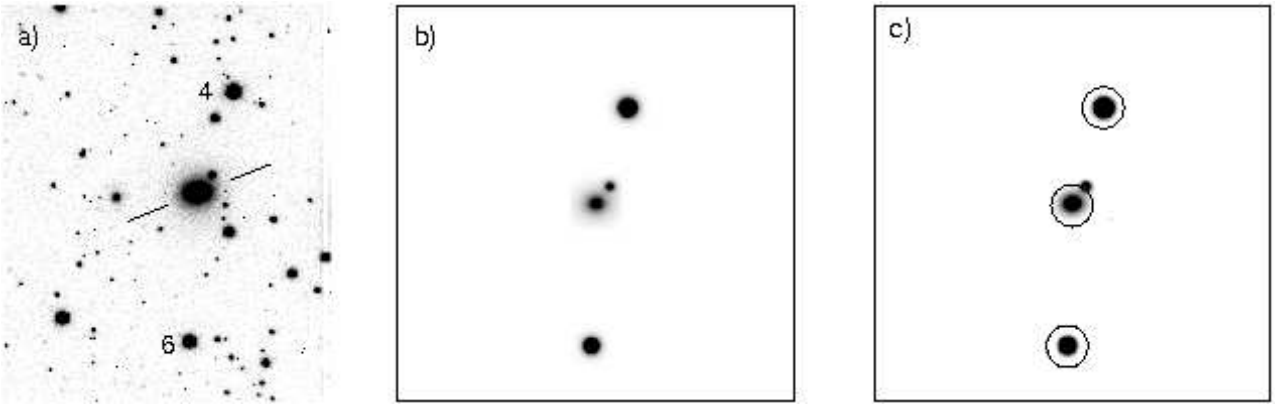}
\caption{\label{esimerkki1} An example of the modeling procedure: a)
An R-band image of 1ES 1959+650 obtained with the NOT with a FWHM of
0\farcs7. The field size is 2.5 $\times$ 3.0 arcmin. The primary
comparison stars 4 and 6 from Villata et al. (\cite{vill98}) have also
been marked.  b) The simulated field, convolved to a FWHM of 3\farcs0
with a $\beta$=2.5 Moffat profile. c) The aperture photometry with an
aperture radius of 10\farcs0. Note that using a large aperture
includes a substantial fraction of the nearby star.  }
\end{figure*}

\subsection{Photometric calibration}

For maximum accuracy, the the host galaxy fits should be calibrated
in exactly the same way as the differential photometry.  Since the
latter is calibrated using comparison stars in the object field,
ideally the deep host galaxy images should contain the same comparison
stars. However, this was the case for three objects only (ON 325, OT 546
and 1ES 1959+650), in the rest the comparison stars were either out of the
field of view or strongly saturated. For these sources we first
selected 1-3 relatively bright (but unsaturated) stars from the deep
images and determined their R-band magnitudes using our $\sim$ 4 years
of monitoring data with the comparison stars in table
\ref{primcompdata} as calibrators. This effectively ties the deep
images to the same scale with our photometry, although the uncertainty
of the host galaxy increases due to the adoption of secondary (and
more noisy) calibrators.

\subsection{Modeling and aperture photometry}

After the model fitting and photometric calibration we created
simulated master fields of all objects in Table \ref{hostidata} (see
Fig. \ref{esimerkki1}).  These fields contained the host galaxy
(without the BL Lac nucleus), any nearby sources and the primary
comparison star(s) in Table \ref{primcompdata}.  The models were made
with the same pixel scale and software as in the model fitting, thus
ensuring accurate representation of the light distribution in the
original images. After this, the master fields were convolved with
three different PSF profiles (Moffat profiles with $\beta$ = 2.0, 2.5
and 3.0) over a grid of FWHM values 0.5, 1.0 \ldots, 8.0 arcsec,
resulting in 48 modeled fields per object.

\begin{figure}[t]
\begin{center}
\includegraphics[width=11cm]{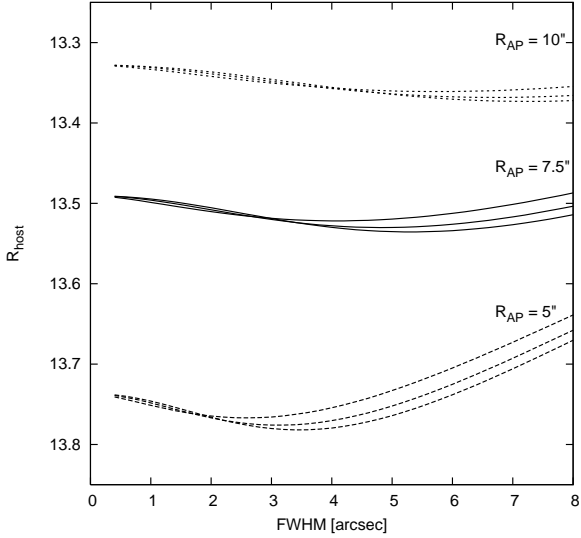}
\end{center}
\caption{\label{mrk501esim}The host galaxy magnitude of Mrk 501
as a function of FWHM and three aperture radii. The three curves for
each aperture correspond to the three different convolution kernels
(Moffat $\beta$ = 2.0, 2.5 and 3.0).}
\end{figure}

The simulated fields were then measured with aperture photometry using
the same procedure and software as for the real monitoring data,
i.e. the primary comparison stars were used as calibrators and the
magnitudes were corrected for the color difference between the object
and the comparison stars using calibrated color coefficients.  We
generally do not know the ($V-R$) color of the host galaxies, thus we
assumed ($V-R$) = 0.6 for them, corresponding to an early-type galaxy
at z=0. As the host galaxies of all our sources are early type
galaxies at relatively low redshifts, and the ($V-R$) color
coefficients relatively small ($<$0.1), this assumption does not
introduce large errors.  For the measurements we used aperture radii of
0.5, 1.0, \ldots, 10.0 arcsec, covering the expected range of aperture
radii in long-term photometric monitoring programs.

\section{Results and discussion}

The R-band host galaxy fluxes resulting from the above procedure are
given in the tables B.2-B.21 in Appendix B (available online only, see
Table B.1 for an example). The fluxes are given as a function of
aperture radius and FWHM (in arcsec) for the $\beta$=2.5 convolution
kernel. Note that we show the data only for aperture radii and FWHM
larger than the resolution of the deep R-band images.  In Figure
\ref{mrk501esim} we show an example of the host galaxy magnitude as a
function of FWHM and three aperture radii. The figure illustrates well
the most important aspects of the data,which we discuss now in more
detail.

\begin{figure*}[t]
\begin{center}
\includegraphics[width=17cm]{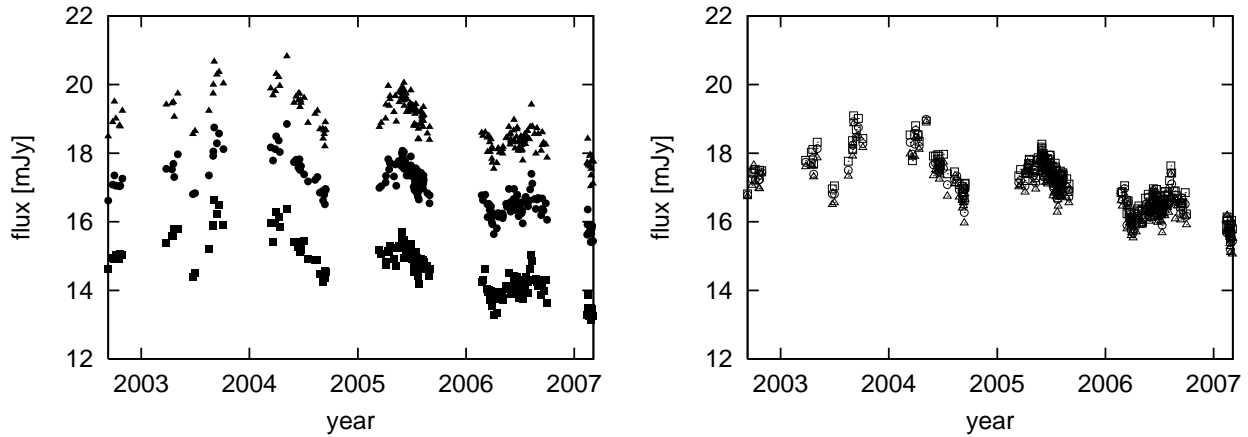}
\end{center}
\caption{\label{vahennys} The light curves of Mrk 501 before (left)
and after (right) reducing all measurements to 7\farcs5 aperture
radius using the host galaxy correction derived in this study. The
curves on the left correspond to (from top to bottom) 10.0, 7.5 and
5.0 arcsec aperture radii. The same symbols have been used in both
panels, except that in the right panel they are unfilled.}
\end{figure*}

As can be expected, the host galaxy brightness depends strongly on the
chosen aperture radius.  The dependence on FWHM is smaller, typically
a few hundreds of a magnitude even for large (factor of two) changes
of FWHM. Note that the curves in Fig. \ref{mrk501esim} are valid for
the host galaxy only. Normally there is a nuclear component present,
which makes the dependence on seeing even smaller. Thus the dependence
on seeing is probably too small to be of importance, with the
exception of microvariability studies where very small amplitude
variations are looked for (Cellone et al. \cite{cell00}).

We next study the possibility to reduce all observations to the same
aperture radius and FWHM using our 4 years of monitoring data as a
test bench.  We measured the light curves of the objects in Table
\ref{hostidata} using three different aperture radii, 5.0, 7.5 and
10.0 arcsec. The median FWHM of our observations is 3.4 arcsec, so the
smallest aperture probably best matches our typical seeing and the
largest would rarely be used in real monitoring work. We nevertheless
consider all three here for an illustrative example.  The FWHM was
determined by fitting $\beta$=2.5 Moffat profiles to the stellar
images in each monitoring frame. The host galaxy magnitude
corresponding to the aperture radius and FWHM was then looked from
tables B.2-B.21, the host-subtracted magnitude computed for each
frame, and all fluxes within one night (typically 3-4) were
averaged.

In Fig. \ref{vahennys} we show one example of the outcome of this
procedure. The left panel shows the light curves of Mrk 501 using the
three aperture radii. The presence of the host galaxy is clearly
revealed by the increasing average object brightness with increasing
aperture radius. This should not happen with point sources since we
are using {\em differential} aperture photometry, unless there are
e.g. large PSF variations across the field of view. The shape of the
light curve is independent of aperture radius, except that the light
curves measured with larger aperture are noisier due to too large
aperture with respect to the seeing.

\begin{figure}[t]
\begin{center}
\includegraphics[width=9cm]{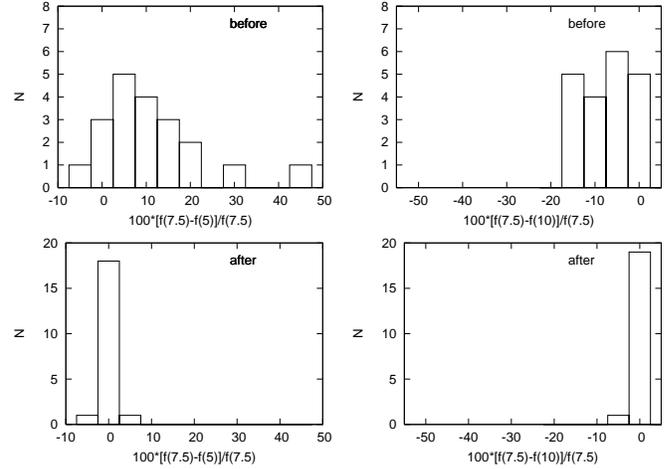}
\end{center}
\caption{\label{jakaumat} The percent difference between 7\farcs5 and
5\arcsec (left) and 7\farcs5 and 10\arcsec (right) light curves before
(upper row) and after (lower row) applying the host galaxy correction
for all 20 sources.  }
\end{figure}

The right panel shows the same light curves after reducing all
measurements to a 7\farcs5 aperture radius. The average levels of the
light curves agree now much better, the difference between average
levels of 7\farcs5 and 5\arcsec curves is now 0.9\%, compared to
13.8\% before the correction. The corresponding values between the
7\farcs5 and 10\arcsec curves are 1.0\% and 11.0\% Similar improvement
is observed in all 20 light curves. This is summarized in
Fig. \ref{jakaumat}, where we show the differences between 7\farcs5
and 5\arcsec and 7\farcs5 and 10\arcsec light curves before and after
the corrections (in percent).  The distributions after the correction
are concentrated around zero with an average of (0.0$\pm$0.4) for the
lower left distribution and (-0.1$\pm$0.3) for the lower right
distribution, showing that on average the correction works very well.
The standard deviations of the distributions are 1.5 and 1.2 for the
lower left and lower right distributions, respectively.  These numbers
represent the accuracy one can expect for these sources when trying to
reduce observations made with different aperture radii into the same
aperture. For instance, if observations are measured with 5.0 and 7.5
arcsec radii, one can expect the measurements to agree within 1-2\%
(depending on the source) after the correction.  

We have also looked at similar distributions as in Fig. \ref{jakaumat}
for pure nuclear fluxes, i.e. after subtraction of the host galaxy
correction. In this case the relative accuracy is lower, simply due to
the fact that pure nuclear fluxes are lower than total fluxes.  The
distributions of the corrected fluxes have now standard deviations of
4.7\% (7.5 vs. 5.0 arcsec aperture) and 4.0\% (7.5 vs. 10.0 arcsec
aperture). The total range is -7.3\% to 13.8 \% (7.5 vs. 5.0 arcsec)
and -6.5\% to 12.1\% (7.5 vs. 10.0 arcsec), showing that differences
over 10\% in nuclear fluxes are possible after correction for the host
galaxy.

The above results emphasize the importance of using the same aperture
radius between different observers in large monitoring campaigns. In
principle the tables in Appendix B could be used to reduce all data
into the same aperture, but in practice the accuracy of this
correction may not be sufficient for all applications.  In large
monitoring campaigns it is thus far better to secure in advance that
all parties are using the same aperture when performing reductions
than try to correct for different aperture sizes afterwards.

We finally mention an example of the application of the tables in
Appendix B.  Our optical monitoring work is mainly done to support the
TeV observations by the MAGIC-telescope on La Palma. In this work it
is important to subtract the host galaxy contribution correctly to
build an accurate SED and to correctly estimate the relative
variability amplitude. Host galaxy subtraction thus forms an important
step before using the data for comparisons with the models. As already
mentioned, the host galaxy flux can be a substantial fraction of the
total flux.  For instance, the average optical flux of 1ES 2344+514
during the four years of our monitoring is 4.43 mJy (using an aperture
radius of 7.5 arcsec) with an rms scatter of 0.1 mJy, i.e. 2.2\% of
the average level.  However, from table B.21 we see, that
the host galaxy flux at 7.5 arcsec aperture radius and 3 arcsec FWHM
(closest to our average FWHM of 3.4 arcsec) is 3.71 mJy, i.e. the true
nuclear flux is only 0.72 mJy and the 0.1 mJy rms variability
corresponds to 14\% of the mean level.  The above example shows how
important it is to properly subtract the host galaxy contribution
before proceeding with the analysis.

\subsection{\label{errorcalc}Error analysis}

The accuracy of our method depends mainly on three factors: 1) how
accurately is the fitting procedure able to separate the nuclear
component from the host galaxy component, 2) how accurately does the
S\'{e}rsic model represent the true surface brightness profile and 3)
how accurately can the deep R-band images be calibrated. 

As noted by several authors before (e.g. Nilsson et al.  \cite{nil03},
Pursimo et al. \cite{pur02}, Scarpa et al. \cite{sca00a}) the
differences between the host galaxy parameters derived by different
authors tend to be significantly higher than the quoted error bars
(see e.g. Fig 3. by Nilsson et al. \cite{nil03}, but note also that
one should check that the same model and same definitions for the
parameters are used when making the comparisons). For instance, the
rms difference in the host galaxy magnitude is typically 0.2-0.3 mag,
but the quoted error bars are typically a factor of 2-5 lower.  The
effective radius is notoriously difficult to constrain, with often
factor of two differences for even well-resolved objects. These
differences point to systematical errors that probably stem from many
factors, such as the nucleus to host galaxy flux ratio (bright nuclei
make it more difficult to constrain the host galaxy), redshift
(distant host galaxies are fainter and smaller) the method of
analysis, resolution, and signal to noise.  For instance, by analyzing
very high signal to noise NOT images Pursimo at al. (\cite{pur02})
found that the derived effective radius depends on the outer radius of
the fit region with larger fit radii producing larger effective
radii. Furthermore, they noted that for the apparently largest
galaxies their values for effective radius were systematically larger
than those obtained from HST data, which was attributed to relatively
short exposures of the HST images.
 
Test simulations in Nilsson et al. \cite{nil03} showed that the host
galaxy and nuclear parameters can be recovered without bias if the PSF
has sufficiently high S/N and the apparent size of the host galaxy is
sufficiently large (i.e. several times larger than the FWHM).  Since
these two conditions are fulfilled here (all our host galaxies are
nearby, z$<$0.4), the procedure is expected to accurately separate the
nuclear component from the host galaxy component and provide accurate
morphological parameters (magnitude, effective radius, ellipticity and
position angle) for the host galaxy.  However, even with well-resolved
images one may encounter problems in characterizing the host
galaxy. Many elliptical galaxies exhibit deviations from a smooth
S\'{e}rsic profile: disk components in their central parts (Scorza \&
Bender \cite{scoben95}), central ``cusps'' (Trujillo et
al. \cite{tru2004}) and even dust lanes (e.g. 1ES 1959+650; Heidt et
al. \cite{heidt99}). If the host galaxy surface brightness does not
follow the S\'{e}rsic law, errors may in induced to the host galaxy
magnitudes.

Considering the points above we have divided the total error into
three components: the random error $\sigma_{\rm ran}$, the
systematic error $\sigma_{\rm sys}$ and the calibration error
$\sigma_{\rm cal}$.  The random error arises from random noise
in the images (photon and readout noise), from the errors in the
assumed PSF shape and background determination errors.  In a perfect
situation, free of systematic effects, this would be the only source
of noise. We created 30-50 simulated images for each source with the
same noise characteristics as in the real data, an error in the
background level and a realistic representation of PSF variations over
the field of view (see Nilsson et al. \cite{nil03} for details). We
then fitted a two-dimensional model to each image using the same
procedures as with the real data. After this, the host galaxy flux was
determined over the same grid of FWHM values and aperture radii as
with the real data.  Finally, the random error, $\sigma_{\rm
ran}$, was computed as the standard deviation of the measured fluxes
at each aperture radius and FWHM. Typically these errors are small (a
few percent), except for very small aperture radii and small FWHM
values and depending on the apparent size of the host galaxy.

As discussed above, systematic effects and deviations from a smooth
S\'{e}rsic law are possible and thus should be included in the error
estimate. Note however, that the errors in host galaxy magnitude and
effective radius are correlated, i.e. any deviation in one parameter
is ``corrected'' by the fitting procedure by adjusting the other to
match the observed 2-d light distribution.  Since we are mainly
interested in the accuracy of the 2-d representation of the source and
not the parameters themselves, the best way to estimate the systematic
errors is to examine the model residuals.  Any deviations from a
S\'{e}rsic law or any systematic effects should be visible as nonzero
residuals and can be quantified by measuring the residual fluxes over
the aperture radii concerned here. The only type of error we cannot
estimate properly is wrong separation of the host galaxy flux from the
nuclear flux. For instance, if the host galaxy has a flattened profile
at distances from the center smaller than our resolution, our model
will overestimate the host galaxy flux and underestimate the nuclear
flux. Due to insufficient resolution this may not be detected in the
residual image.

Many objects in our sample show deviations from a smooth S\'{e}rsic
profile e.g. due to interaction with a nearby companion galaxy or dust
lanes. These deviations are usually not very large (from a few percent
up to 20\%) and they tend to be in the form of fluctuations around
zero mean and thus mostly cancel out, especially when large aperture
radii are used. Some sources (like RGB J1117+202; see
Fig. \ref{profiilit}) show larger deviations (up to 50\%) in their
outer parts. Since this happens at large distances from the object
center and at low surface brightnesses, it has a very small effect in
aperture magnitudes.

Hence, the systematic effects ($\sigma_{\rm sys}$) were estimated in
the following way: the residual images (observed - model) were first
convolved with $\beta = 2.5$ Moffat profiles to FWHM values 1\farcs0,
2\farcs0,...,8\farcs0 and these images were measured with aperture
radii 1\farcs0, 1\farcs5,...,10\farcs0. The norm of the measured flux
at each FWHM and aperture radius was then formally adopted as the
systematic error $\sigma_{\rm sys}$, giving us a very conservative
estimate of the systematic error.

After computing $\sigma_{\rm ran}$ and $\sigma_{\rm sys}$ we compute
the final error $\sigma$ at each aperture radius and FWHM from
$$
\sigma = \sqrt{\sigma_{\rm ran}^2 + \sigma_{\rm sys}^2 +
\sigma_{\rm cal}^2}\ ,
$$
where $\sigma_{\rm cal}$ is the estimated error of the calibration
(0.01-0.05 mag). Note that contrary to the tables in Appendix B the
error bars in Table \ref{hostidata} contain the contributions from
$\sigma_{ran}$ and $\sigma_{cal}$ only.

As a general remark we note that the relative error tends to be
largest at small aperture radii and small FWHM values. This is
understandable since the central regions are most sensitive to errors
in the model parameters (large $\sigma_{\rm ran}$) and in the central
parts deviations from a S\'{e}rsic profile and PSF error are also
largest (large $\sigma_{\rm sys})$. 

\subsection{Future work}

In the future this method can be easily extended to more objects and
other wavelength bands. The only requirement is a high-resolution,
high signal to noise image of the object at the given band. An
important step would be a more detailed study of the influence of the
PSF shape, presently assumed to be a Moffat $\beta=2.5$ profile.  In
principle, the PSF could be determined from the image using a
sufficiently bright star, the photometric model convolved with this
PSF and the aperture correction computed for each frame individually,
yielding a more accurate correction. This could potentially improve
times series of objects with bright host galaxies. However, poor
sampling and the fact that such stars are not always available could
mean that this method is not always usable.

The method could also be used in imaging polarimetry, where dilution
by the unpolarized host galaxy light makes it very difficult to
estimate the true polarization of the BL Lac nucleus. Since many
polarization techniques use aperture photometry, a similar technique
than employed here could be used to overcome the dilution. Finally,
the accuracy of the method at small aperture radii and small FWHM
could be improved using HST images, since these allow a better
characterization of the central parts of the host galaxies.

\subsection{Notes on individual objects}

{\bf 1ES 0033+595}: There is a star 1.6 arcsec away from this object
with a comparable magnitude (R = 17.9) to the object (R $\sim$ 17.1).
This pair is unresolved in typical monitoring observations, and in
practice one cannot center the aperture on 1ES 0033+595, unless the
seeing is very good ($<$ 1 arcsec).  In our simulations we have thus
fixed the aperture center midway between 1ES 0033+595 and the star to
make a more realistic simulation.

\noindent
{\bf Mrk 421} Only one very faint star (R = 17.5) is available in the
field as a PSF star and a secondary calibrator. Thus the errors for
Mrk 421 are larger than for the other well-resolved objects.

\noindent
{\bf RGB J1117+202} Our model underestimates the host galaxy brightness
at large radii. It is possible that the galaxy is interacting with the
two nearby galaxies thus distorting its shape, although no clear signs
of interaction are visible. This affects very large apertures ($\ga 8$
arcsec) only.

\noindent
{\bf 1ES 1959+650} There is a weak dust lane in this galaxy roughly 1
arcsec North of the nucleus oriented in E-W direction (Heidt et
al. \cite{heidt99}). This dust lane is not included in our model,
but its effect is very small ($<$ 0.01 mag) judging from the photometry
on the residual image.

\noindent
{\bf BL Lac} The light from a bright (R = 11.93) star 24 arcsec E of
this object enters the aperture at large seeing values and large
apertures.  Since this effect depends strongly on the PSF shape
at the outer parts, we have not included it in our model, but taken it
into account in our error estimates.

\section{Conclusions}

Photometric monitoring of active galactic nuclei is often complicated
by the presence of a strong host galaxy component.  The host galaxy
distorts the optical fluxes and thus makes it difficult to estimate
the SED and true variability level of the active nucleus. In
addition, FWHM changes can induce false variability, which complicates
e.g. microvariability studies.

In order to quantify these effects, we have measured the host galaxy
flux for 20 BL Lacertae objects over a grid of aperture radii and FWHM
values using high-resolution images obtained at the Nordic Optical
Telescope (NOT). In addition to the host galaxy, we have also included
the flux from any nearby stars or galaxies that could contribute
significantly inside the measurement aperture. Two-dimensional model
fitting was employed to separate the nuclear component from the host
galaxy component and to determine the photometric parameters of the
host galaxy. Using these parameters the host galaxy flux as a function
aperture radius and FWHM was determined assuming a Moffat $\beta =
2.5$ profile for the PSF. We have also estimated the error bars taking
into account fitting errors, possible deviations from the assumed
S\'{e}rsic profile and calibration errors.

The results are given in Tables B.2-B.21 in Appendix B (available
online only, see Table \ref{fesimerkki} for an example), which give
the integrated fluxes of ``contaminating'' sources (host galaxy and
significant nearby sources) as a function of aperture radius and
FWHM. We found that the host galaxy flux depends quite strongly on the
aperture radius, but FWHM usually has a minor effect (a few percent).
We have tested the correction tables using our 4 years of monitoring
data obtained at small (0.35 to 1m) telescopes. We found the
correction to work very well on average, typically we can reduce
observations obtained with different aperture radii to a common
aperture with an accuracy of $\sim$ 1-2\%. For pure nuclear fluxes the
accuracy is $\sim$ 5\% on average, but can be over 10\% in some cases.

It is important to note a few caveats with respect to the tables in
Appendix B. Firstly, the results are calibrated using the comparison
star data in Table \ref{primcompdata}. If a different calibration
scheme is adopted, the corresponding corrections to Tables in Appendix
B has to be computed.  Secondly, it is possible that in some sources
the two dimensional fitting routine has failed to correctly separate
the nuclear flux from the host galaxy flux due to deviations from a
S\'{e}rsic law in the inner part of the host galaxy. Since this effect
is very difficult to quantify, it has not been included in the error
estimates. And thirdly, throughout this work we have assumed a Moffat
$\beta = 2.5$ profile for the PSF. Real data typically present a wide
variety of PSF shapes and thus our correction is not an accurate
estimate in all cases. However, the tests with actual monitoring data
presented here show that on average the correction works very well and
it can be reliably used to subtract the host galaxy contribution, so
the above effects do not change our results in any significant way.

\begin{acknowledgements}
The authors thank R. Falomo for providing deep R-band images for three
objects.  These data are partly based on observations made with the
Nordic Optical Telescope, operated on the island of La Palma jointly
by Denmark, Finland, Iceland, Norway, and Sweden, in the Spanish
Observatorio del Roque de los Muchachos of the Instituto de
Astrofisica de Canarias.  Part of the data presented here have been
taken using ALFOSC, which is owned by the Instituto de Astrofisica de
Andalucia (IAA) and operated at the Nordic Optical Telescope under
agreement between IAA and the NBIfAFG of the Astronomical Observatory
of Copenhagen.
\end{acknowledgements}

\clearpage

\begin{appendix}

\section{New comparison star sequences}

\begin{figure*}
\includegraphics[width=5.8cm]{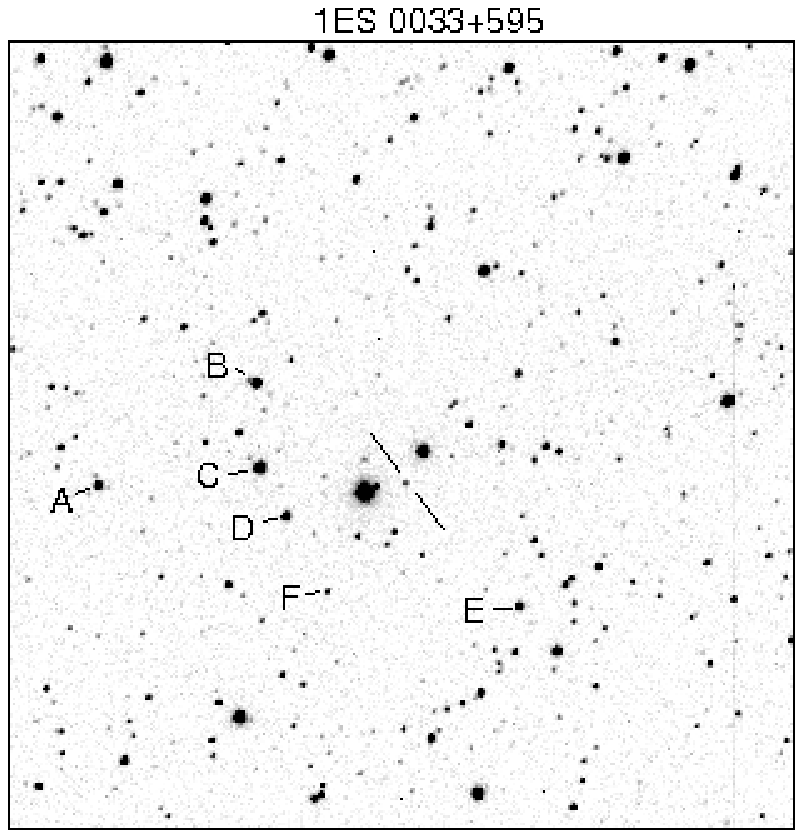}
\includegraphics[width=5.8cm]{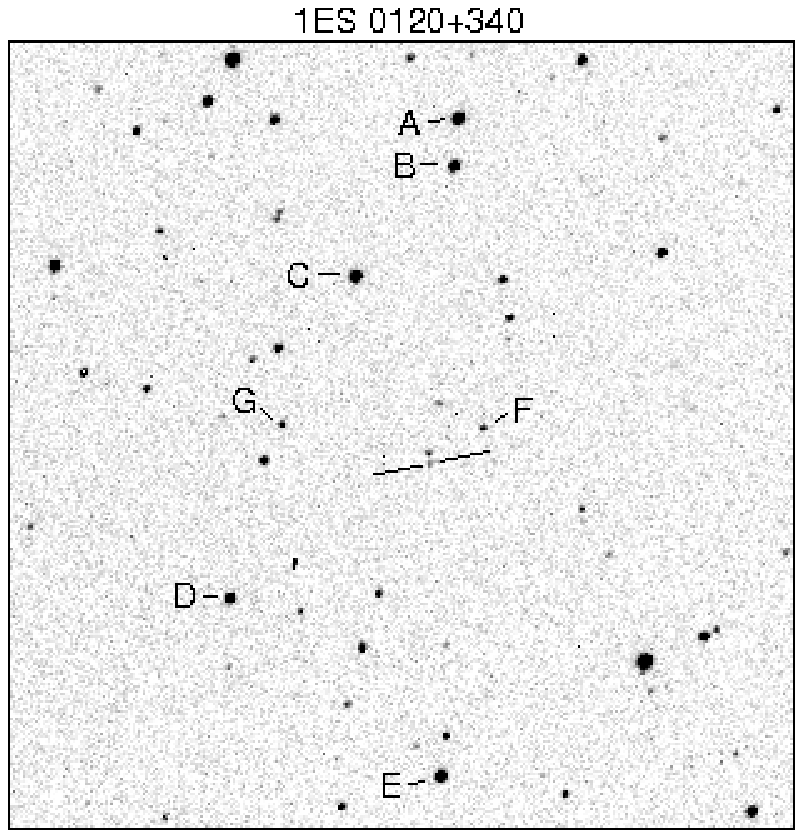}
\includegraphics[width=5.8cm]{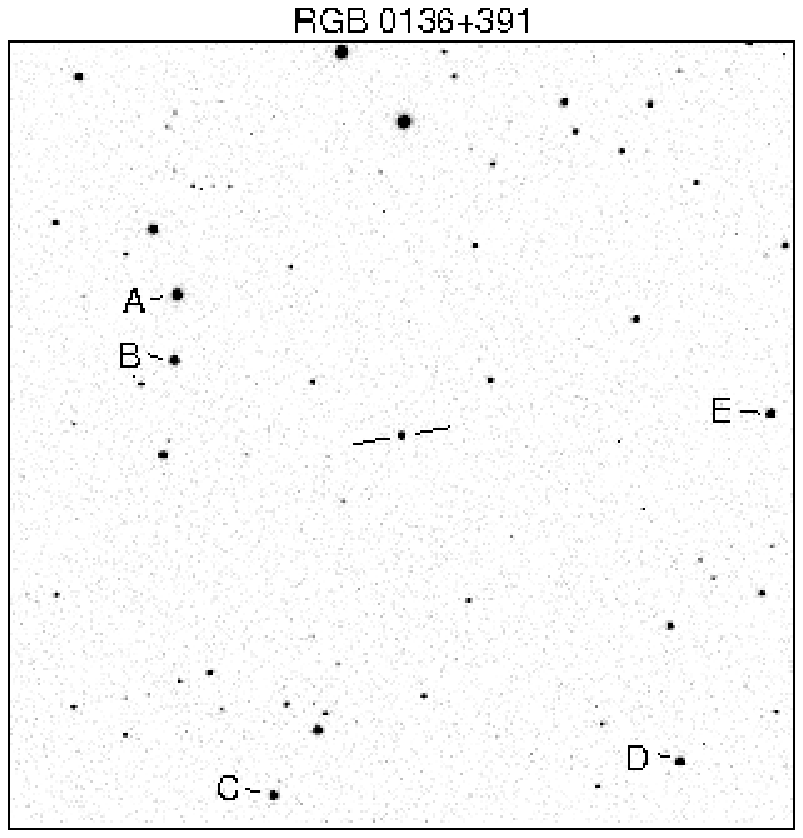}\\

\includegraphics[width=5.8cm]{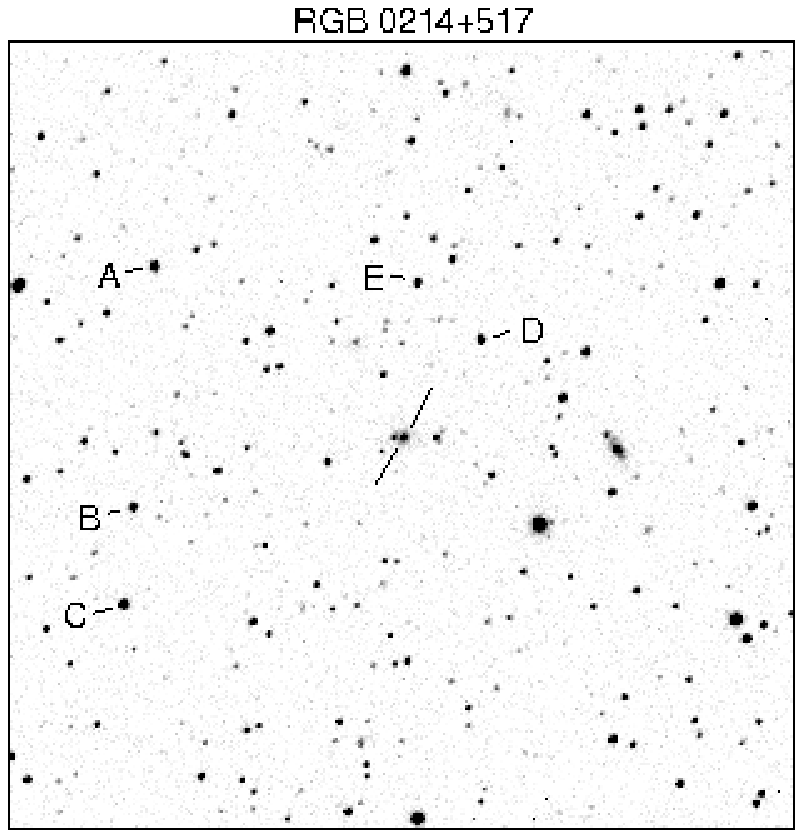}
\includegraphics[width=5.8cm]{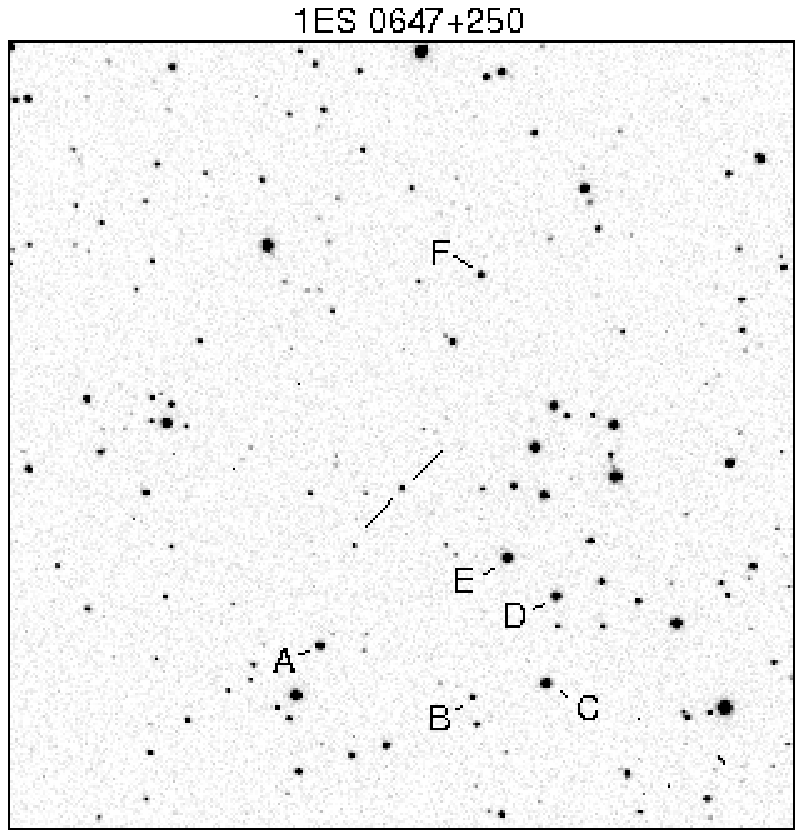}
\includegraphics[width=5.8cm]{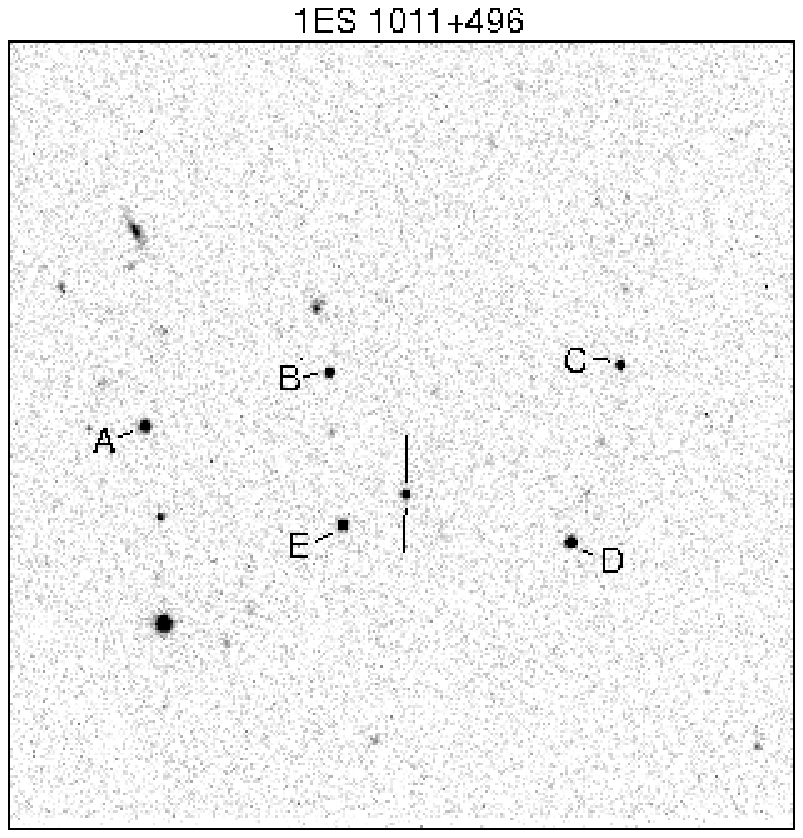}\\

\includegraphics[width=5.8cm]{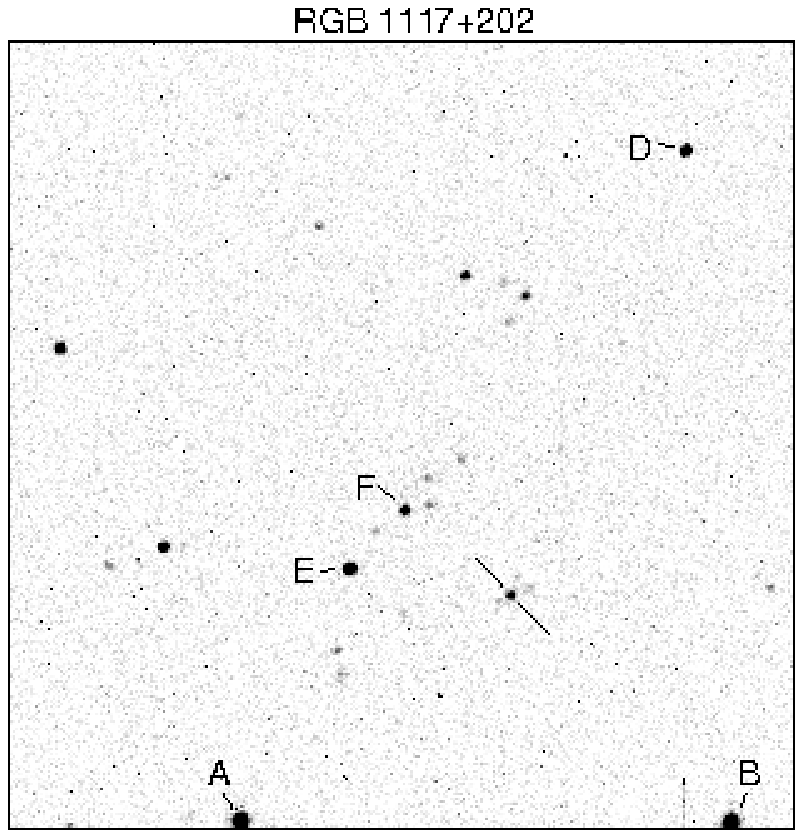}
\includegraphics[width=5.8cm]{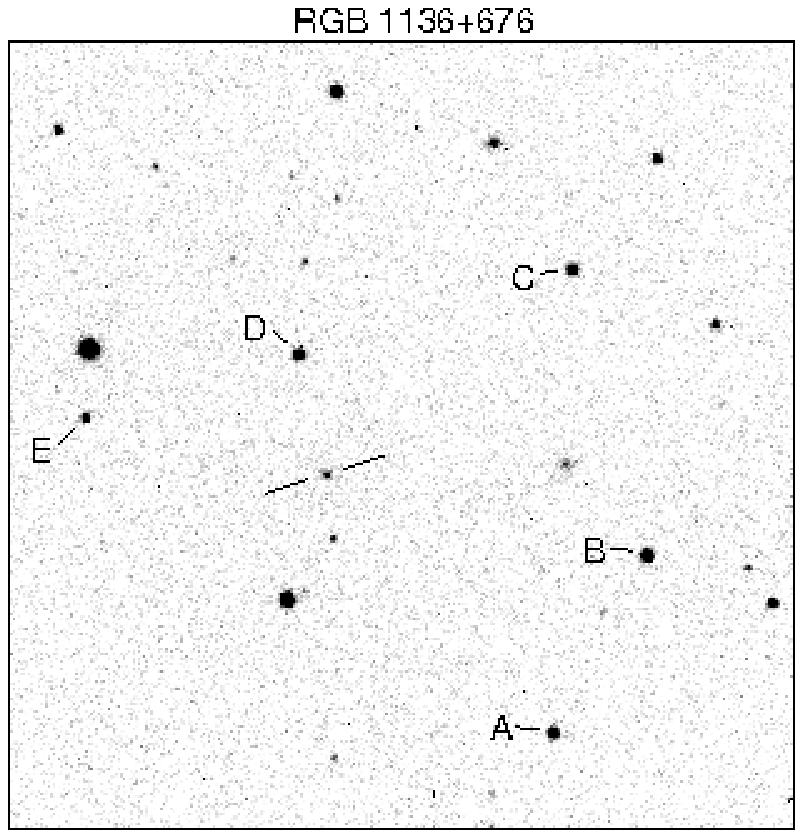}
\includegraphics[width=5.8cm]{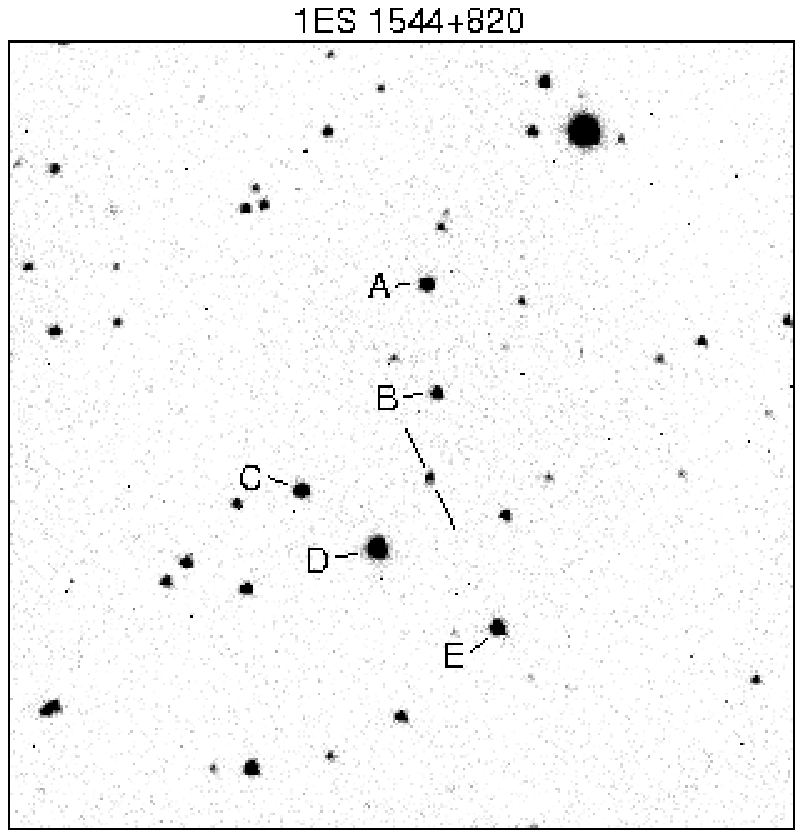}
\caption{\label{fcharts}Finding charts for the new comparison star sequences.
North is up and east is to the left in all images. The field size is 8 arcmin.}
\end{figure*}

We have calibrated new comparison star sequences for previously
uncalibrated fields using the 1.03 m telescope at Tuorla Observatory
during 7 photometric nights in 2003-05. The observations were made
through V- and (Cousins) R-band filters using a Santa Barbara
ST-1001E CCD-camera with a gain factor of 3.1 $e^-$ ADU$^{-1}$ and a
readout noise of 17~$e^-$. Finding charts of the new comparison stars
can be found in Fig. \ref{fcharts}.

During each night several BL Lac fields were observed together with
10-15 bright northern photometric standard stars from Cousins
(\cite{cous84}), Jerzykiewicz \& Serkowski (\cite{jerzy66}), Landolt
(\cite{lana}, \cite{lanb}), Oja (\cite{oja}), Taylor (\cite{taylor}),
Taylor \& Joner (\cite{tay85}) and Taylor et al. (\cite{tay89}).  The
observed frames were bias- and dark-subtracted and flat-fielded in a
standard manner and instrumental magnitudes of the BL Lac comparison
stars and the standard stars were determined using aperture
photometry.

To determine the transformation from instrumental to standard
magnitudes we fitted the standard star observations with the formulae
$$
R = r + \zeta_R + k_R X + \epsilon(V-R)
$$
$$
V - R = \zeta_{VR} + k_{VR} X + \psi(v-r)\ ,
$$
where V and R are standard magnitudes, v and r instrumental
magnitudes, $\zeta$ is the zero point, $k$ the first-order extinction
coefficient, $\epsilon$ and $\psi$ the color terms and $X$ the
airmass. The range of values found for $k_R$ and $k_{VR}$ were
0.10-0.15 and 0.05-0.09, respectively. The color terms were found to
be stable over the whole observing period with average and rms scatter
of $0.050 \pm 0.011$ ($\epsilon$) and $0.77 \pm 0.03$ ($\psi$).  After
the fitting the comparison star magnitudes were transformed to the
standard system using the adopted transformation constants. Table
\ref{compmagnitudes} gives the resulting magnitudes.

\begin{table}
\caption{\label{compmagnitudes}Comparison star magnitudes. The third column gives the number of
observations for each star.}
\centering
\begin{tabular}{lllll}
\hline\hline
Object       & Star & N$_{\rm obs}$ & R & V-R\\
\hline
1ES 0033+595 & A  & 3  &  14.10 $\pm$ 0.03 &  1.06 $\pm$ 0.04\\
1ES 0033+595 & B  & 3  &  13.33 $\pm$ 0.02 &  0.56 $\pm$ 0.03\\
1ES 0033+595 & C  & 2  &  12.52 $\pm$ 0.03 &  0.74 $\pm$ 0.04\\
1ES 0033+595 & D  & 3  &  13.66 $\pm$ 0.03 &  1.46 $\pm$ 0.04\\
1ES 0033+595 & E  & 3  &  13.91 $\pm$ 0.02 &  0.55 $\pm$ 0.04\\
1ES 0033+595 & F  & 3  &  16.67 $\pm$ 0.03 &  0.87 $\pm$ 0.07\\
\\
1ES 0120+340 & A  & 2  &  13.13 $\pm$ 0.03 &  0.50 $\pm$ 0.05\\
1ES 0120+340 & B  & 2  &  13.74 $\pm$ 0.03 &  0.48 $\pm$ 0.05\\
1ES 0120+340 & C  & 2  &  13.12 $\pm$ 0.03 &  0.38 $\pm$ 0.05\\
1ES 0120+340 & D  & 2  &  14.02 $\pm$ 0.03 &  0.51 $\pm$ 0.05\\
1ES 0120+340 & E  & 2  &  13.55 $\pm$ 0.03 &  0.56 $\pm$ 0.05\\
1ES 0120+340 & F  & 2  &  16.76 $\pm$ 0.04 &  0.57 $\pm$ 0.07\\
1ES 0120+340 & G  & 1  &  16.43 $\pm$ 0.04 &  0.45 $\pm$ 0.08\\
\\
RGB J0136+391 & A  & 3  &  13.13 $\pm$ 0.02 &  0.60 $\pm$ 0.04\\
RGB J0136+391 & B  & 3  &  13.82 $\pm$ 0.02 &  0.42 $\pm$ 0.04\\
RGB J0136+391 & C  & 3  &  14.40 $\pm$ 0.03 &  0.76 $\pm$ 0.04\\
RGB J0136+391 & D  & 3  &  14.42 $\pm$ 0.03 &  0.46 $\pm$ 0.04\\
RGB J0136+391 & E  & 3  &  14.84 $\pm$ 0.03 &  0.51 $\pm$ 0.04\\
\\
RGB J0214+517 & A  & 1  &  13.85 $\pm$ 0.05 &  0.51 $\pm$ 0.06\\
RGB J0214+517 & B  & 1  &  14.54 $\pm$ 0.05 &  0.57 $\pm$ 0.06\\
RGB J0214+517 & C  & 1  &  13.61 $\pm$ 0.05 &  0.38 $\pm$ 0.06\\
RGB J0214+517 & D  & 1  &  15.09 $\pm$ 0.05 &  0.47 $\pm$ 0.07\\
RGB J0214+517 & E  & 1  &  15.19 $\pm$ 0.05 &  0.44 $\pm$ 0.07\\
\\
1ES 0647+250 & A  & 1  &  13.83 $\pm$ 0.04 &  0.33 $\pm$ 0.05\\
1ES 0647+250 & B  & 1  &  15.22 $\pm$ 0.04 &  0.38 $\pm$ 0.06\\
1ES 0647+250 & C  & 1  &  13.40 $\pm$ 0.04 &  0.40 $\pm$ 0.05\\
1ES 0647+250 & D  & 1  &  13.44 $\pm$ 0.04 &  0.34 $\pm$ 0.05\\
1ES 0647+250 & E  & 1  &  13.03 $\pm$ 0.04 &  0.59 $\pm$ 0.05\\
1ES 0647+250 & F  & 1  &  14.89 $\pm$ 0.04 &  0.38 $\pm$ 0.05\\
\\
1ES 1011+496 & A  & 3  &  13.40 $\pm$ 0.03 & 0.47 $\pm$ 0.03\\
1ES 1011+496 & B  & 3  &  15.44 $\pm$ 0.03 & 0.44 $\pm$ 0.04\\
1ES 1011+496 & C  & 3  &  15.42 $\pm$ 0.03 & 0.31 $\pm$ 0.04\\
1ES 1011+496 & D  & 3  &  14.01 $\pm$ 0.03 & 0.31 $\pm$ 0.03\\
1ES 1011+496 & E  & 3  &  14.04 $\pm$ 0.03 & 0.39 $\pm$ 0.03\\
\\
RGB J1117+202 & A  & 2  &  11.90 $\pm$ 0.04 & 0.38 $\pm$ 0.04\\
RGB J1117+202 & B  & 2  &  12.02 $\pm$ 0.04 & 0.38 $\pm$ 0.04\\
RGB J1117+202 & D  & 2  &  14.82 $\pm$ 0.04 & 0.70 $\pm$ 0.05\\
RGB J1117+202 & E  & 2  &  13.56 $\pm$ 0.04 & 0.42 $\pm$ 0.04\\
RGB J1117+202 & F  & 2  &  15.16 $\pm$ 0.04 & 0.43 $\pm$ 0.05\\
\\
RGB J1136+676 & A  & 1  &  14.48 $\pm$ 0.04 & 0.66 $\pm$ 0.06\\
RGB J1136+676 & B  & 1  &  14.22 $\pm$ 0.04 & 0.48 $\pm$ 0.05\\
RGB J1136+676 & C  & 1  &  14.73 $\pm$ 0.04 & 0.34 $\pm$ 0.05\\
RGB J1136+676 & D  & 1  &  14.58 $\pm$ 0.04 & 0.46 $\pm$ 0.05\\
RGB J1136+676 & E  & 1  &  15.80 $\pm$ 0.04 & 0.42 $\pm$ 0.07\\
\\
1ES 1544+820 & A  & 3  &  14.59 $\pm$ 0.03 & 0.37 $\pm$ 0.04\\
1ES 1544+820 & B  & 3  &  15.35 $\pm$ 0.03 & 0.80 $\pm$ 0.05\\
1ES 1544+820 & C  & 3  &  14.41 $\pm$ 0.03 & 0.38 $\pm$ 0.04\\
1ES 1544+820 & D  & 3  &  12.87 $\pm$ 0.03 & 0.31 $\pm$ 0.04\\
1ES 1544+820 & E  & 3  &  14.24 $\pm$ 0.03 & 0.41 $\pm$ 0.04\\
\hline
\end{tabular}
\end{table}

To check if any of the comparison stars are variable, we have analyzed
our 4 years of differential photometry of all stars in Table
\ref{compmagnitudes}.  Our initial analysis showed that the
brightnesses of all comparison stars in Table \ref{compmagnitudes}
appear to be constant over the whole 4 year period with a average
scatter of 1.6\% (min. 0.6\%, max 3.6\%) around the mean level.  To
test the variability further we used the $\chi^2$ statistic
$$
\chi^2 = \sum_{i=0}^N \frac{(\overline{x}_i - \mu)^2}{\sigma_i^2}\ ,
$$ 
where $\mu$ is the average flux of the star, $\overline{x}_i$ is the 
weighted average of each star during one night
$$
\overline{x}_i = \frac{\sum_j x_j/\sigma_j^2}{\sum_j 1/\sigma_j^2}
$$
and the corresponding error is
$$
\sigma_i^2 = \frac{1}{\sum_j 1/\sigma_j^2}\ .
$$
In the above summations $j$ runs over one observing night and $i$ over
all observing nights.  The error bars $\sigma_j$ were computed by
taking into account three terms: photon noise of the object, sky noise
in the measurement aperture and background determination error. We
further added 1.5\% in square to each $\sigma_i$ to take into account
flat-fielding errors caused by scattered light in the telescope.

Our null hypothesis is that the flux is constant and we formally
define a star variable if the null hypothesis can be rejected with $p
< 0.1$\% using the above statistic. Computing the $\chi^2$ and its
significance for each star we found that none of the stars shows
significant variability. Taking into account our typical error bars we
can thus conclude that any variability of these stars must
be below 2-4\% level, the exact value depending on the brightness of
the star.

\end{appendix}

\clearpage

\begin{appendix}

\section{Host galaxy fluxes}

The tables here (available online only) give the total contaminating
fluxes (host galaxy + nearby companions) in mJy as a function of
aperture radius and FWHM. The R-band magnitudes have been converted to
linear fluxes $F$ (Jy) using the formula $F=3080.0*10^{-0.4*R}$. Table
\ref{fesimerkki} gives an example.  Note that data is shown only for
aperture radii and FWHM larger than the resolution of the deep R-band
images. The values have not been corrected for galactic
absorption and no K-correction has been applied. The procedure used to
derive the error bars (1$\sigma$) is explained in Section
\ref{errorcalc}.

\onecolumn

\small

%\begin{landscape}
\begin{longtable}[b]{r|cccccccc}
\caption{\label{fesimerkki}The contaminating fluxes (host galaxy + nearby 
companions) of Mrk 501 in mJy. See table \ref{hostidata} for details on the
photometric components included in the model.}\\
\hline\hline
Aperture & \multicolumn{8}{c}{FWHM}\\
radius   & \multicolumn{8}{c}{(arcsec)}\\
(arcsec) &\\
     &  1.0        &2.0        &3.0        &4.0        &5.0        &6.0        &7.0        &8.0\\
\hline
\endfirsthead
 1.0 &  3.4 $\pm$ 0.2 & 4.1 $\pm$ 0.2 & 5.1 $\pm$ 0.2 & 6.0 $\pm$ 0.2 & 6.8 $\pm$ 0.2 & 7.5 $\pm$ 0.2 & 8.2 $\pm$ 0.2 & 8.9 $\pm$ 0.2\\ 
 1.5 &  4.4 $\pm$ 0.2 & 4.8 $\pm$ 0.2 & 5.5 $\pm$ 0.2 & 6.3 $\pm$ 0.2 & 7.0 $\pm$ 0.2 & 7.7 $\pm$ 0.2 & 8.4 $\pm$ 0.2 & 9.0 $\pm$ 0.2\\ 
 2.0 &  5.3 $\pm$ 0.2 & 5.5 $\pm$ 0.2 & 6.0 $\pm$ 0.2 & 6.6 $\pm$ 0.2 & 7.3 $\pm$ 0.2 & 7.9 $\pm$ 0.2 & 8.6 $\pm$ 0.2 & 9.2 $\pm$ 0.2\\ 
 2.5 &  6.2 $\pm$ 0.2 & 6.2 $\pm$ 0.2 & 6.6 $\pm$ 0.2 & 7.1 $\pm$ 0.2 & 7.6 $\pm$ 0.2 & 8.2 $\pm$ 0.2 & 8.8 $\pm$ 0.2 & 9.3 $\pm$ 0.2\\ 
 3.0 &  7.1 $\pm$ 0.2 & 7.0 $\pm$ 0.2 & 7.2 $\pm$ 0.2 & 7.5 $\pm$ 0.2 & 8.0 $\pm$ 0.2 & 8.5 $\pm$ 0.2 & 9.0 $\pm$ 0.2 & 9.5 $\pm$ 0.2\\ 
 3.5 &  7.8 $\pm$ 0.2 & 7.7 $\pm$ 0.2 & 7.8 $\pm$ 0.2 & 8.0 $\pm$ 0.2 & 8.4 $\pm$ 0.2 & 8.8 $\pm$ 0.2 & 9.3 $\pm$ 0.2 & 9.8 $\pm$ 0.2\\ 
 4.0 &  8.5 $\pm$ 0.2 & 8.4 $\pm$ 0.2 & 8.4 $\pm$ 0.2 & 8.5 $\pm$ 0.2 & 8.8 $\pm$ 0.2 & 9.2 $\pm$ 0.2 & 9.6 $\pm$ 0.2 &10.0 $\pm$ 0.3\\ 
 4.5 &  9.2 $\pm$ 0.2 & 9.0 $\pm$ 0.2 & 8.9 $\pm$ 0.2 & 9.1 $\pm$ 0.2 & 9.3 $\pm$ 0.2 & 9.6 $\pm$ 0.2 & 9.9 $\pm$ 0.3 &10.3 $\pm$ 0.3\\
\hline
\end{longtable}
%\end{landscape}

\normalsize

\end{appendix}

\Online

\begin{appendix}

\onecolumn 

\small

\setcounter{table}{1}

\begin{longtable}{c|ccccccc}
\caption{The contaminating fluxes (host galaxy + nearby companions) of 1ES 0033+595 in mJy.}\\
\hline\hline
Aperture & \multicolumn{7}{c}{FWHM}\\
radius   & \multicolumn{7}{c}{(arcsec)}\\
(arcsec) &\\
             &2.0          &3.0          &4.0          &5.0          &6.0          &7.0          &8.0\\
\hline
\endfirsthead
 1.5 & 0.19 $\pm$ 0.01 & 0.19 $\pm$ 0.01 & 0.20 $\pm$ 0.01 & 0.20 $\pm$ 0.02 & 0.21 $\pm$ 0.02 & 0.21 $\pm$ 0.02 & 0.21 $\pm$ 0.03\\ 
 2.0 & 0.20 $\pm$ 0.01 & 0.20 $\pm$ 0.01 & 0.20 $\pm$ 0.02 & 0.21 $\pm$ 0.02 & 0.21 $\pm$ 0.02 & 0.21 $\pm$ 0.02 & 0.21 $\pm$ 0.03\\ 
 2.5 & 0.21 $\pm$ 0.01 & 0.21 $\pm$ 0.01 & 0.21 $\pm$ 0.02 & 0.21 $\pm$ 0.02 & 0.21 $\pm$ 0.02 & 0.21 $\pm$ 0.02 & 0.21 $\pm$ 0.03\\ 
 3.0 & 0.21 $\pm$ 0.02 & 0.21 $\pm$ 0.02 & 0.21 $\pm$ 0.02 & 0.21 $\pm$ 0.02 & 0.21 $\pm$ 0.02 & 0.21 $\pm$ 0.03 & 0.21 $\pm$ 0.03\\ 
 3.5 & 0.21 $\pm$ 0.02 & 0.21 $\pm$ 0.02 & 0.21 $\pm$ 0.02 & 0.21 $\pm$ 0.02 & 0.21 $\pm$ 0.02 & 0.21 $\pm$ 0.03 & 0.21 $\pm$ 0.03\\ 
 4.0 & 0.21 $\pm$ 0.02 & 0.21 $\pm$ 0.02 & 0.21 $\pm$ 0.02 & 0.21 $\pm$ 0.02 & 0.21 $\pm$ 0.03 & 0.21 $\pm$ 0.03 & 0.21 $\pm$ 0.03\\ 
 4.5 & 0.22 $\pm$ 0.02 & 0.21 $\pm$ 0.02 & 0.21 $\pm$ 0.03 & 0.21 $\pm$ 0.03 & 0.21 $\pm$ 0.03 & 0.21 $\pm$ 0.03 & 0.21 $\pm$ 0.03\\ 
 5.0 & 0.22 $\pm$ 0.03 & 0.22 $\pm$ 0.03 & 0.21 $\pm$ 0.03 & 0.21 $\pm$ 0.03 & 0.21 $\pm$ 0.03 & 0.21 $\pm$ 0.03 & 0.21 $\pm$ 0.03\\
\hline
\end{longtable}

\begin{longtable}{c|ccccccc}
\caption{The contaminating fluxes (host galaxy + nearby companions) of 1ES 0120+340 in mJy.}\\
\hline\hline
Aperture & \multicolumn{7}{c}{FWHM}\\
radius   & \multicolumn{7}{c}{(arcsec)}\\
(arcsec) &\\
      & 2.0         & 3.0          &4.0          &5.0          &6.0          &7.0          &8.0\\
\hline
\endfirsthead
 1.5 & 0.10 $\pm$ 0.01 & 0.11 $\pm$ 0.01 & 0.13 $\pm$ 0.01 & 0.15 $\pm$ 0.01 & 0.18 $\pm$ 0.01 & 0.21 $\pm$ 0.01 & 0.24 $\pm$ 0.01\\ 
 2.0 & 0.11 $\pm$ 0.01 & 0.12 $\pm$ 0.01 & 0.14 $\pm$ 0.01 & 0.16 $\pm$ 0.01 & 0.19 $\pm$ 0.01 & 0.22 $\pm$ 0.01 & 0.25 $\pm$ 0.01\\ 
 2.5 & 0.12 $\pm$ 0.01 & 0.13 $\pm$ 0.01 & 0.14 $\pm$ 0.01 & 0.17 $\pm$ 0.01 & 0.20 $\pm$ 0.01 & 0.23 $\pm$ 0.01 & 0.26 $\pm$ 0.01\\ 
 3.0 & 0.13 $\pm$ 0.01 & 0.14 $\pm$ 0.01 & 0.16 $\pm$ 0.01 & 0.18 $\pm$ 0.01 & 0.21 $\pm$ 0.01 & 0.24 $\pm$ 0.01 & 0.27 $\pm$ 0.01\\ 
 3.5 & 0.14 $\pm$ 0.01 & 0.15 $\pm$ 0.01 & 0.17 $\pm$ 0.01 & 0.20 $\pm$ 0.01 & 0.22 $\pm$ 0.01 & 0.25 $\pm$ 0.01 & 0.28 $\pm$ 0.01\\ 
 4.0 & 0.15 $\pm$ 0.01 & 0.17 $\pm$ 0.01 & 0.19 $\pm$ 0.01 & 0.22 $\pm$ 0.01 & 0.24 $\pm$ 0.01 & 0.27 $\pm$ 0.01 & 0.30 $\pm$ 0.01\\ 
 4.5 & 0.17 $\pm$ 0.01 & 0.19 $\pm$ 0.01 & 0.21 $\pm$ 0.01 & 0.24 $\pm$ 0.01 & 0.26 $\pm$ 0.01 & 0.29 $\pm$ 0.01 & 0.31 $\pm$ 0.01\\ 
 5.0 & 0.19 $\pm$ 0.01 & 0.22 $\pm$ 0.01 & 0.24 $\pm$ 0.01 & 0.27 $\pm$ 0.01 & 0.29 $\pm$ 0.01 & 0.31 $\pm$ 0.01 & 0.33 $\pm$ 0.01\\
\hline
\end{longtable}

\clearpage

\begin{longtable}{c|cccccccc}
\caption{The contaminating fluxes (host galaxy + nearby companions) of RGB J0214+517 in mJy.}\\
\hline\hline
Aperture & \multicolumn{8}{c}{FWHM}\\
radius   & \multicolumn{8}{c}{(arcsec)}\\
(arcsec) &\\
      & 1.0    & 2.0         & 3.0          &4.0          &5.0          &6.0          &7.0          &8.0\\
\hline
\endfirsthead
 1.0 & 0.72 $\pm$ 0.03 & 0.89 $\pm$ 0.03 & 1.09 $\pm$ 0.04 & 1.29 $\pm$ 0.04 & 1.47 $\pm$ 0.05 & 1.65 $\pm$ 0.05 & 1.82 $\pm$ 0.06 & 1.97 $\pm$ 0.06\\ 
 1.5 & 0.94 $\pm$ 0.04 & 1.03 $\pm$ 0.04 & 1.18 $\pm$ 0.04 & 1.35 $\pm$ 0.05 & 1.52 $\pm$ 0.05 & 1.69 $\pm$ 0.05 & 1.85 $\pm$ 0.06 & 2.01 $\pm$ 0.06\\ 
 2.0 & 1.15 $\pm$ 0.04 & 1.18 $\pm$ 0.04 & 1.29 $\pm$ 0.04 & 1.44 $\pm$ 0.05 & 1.59 $\pm$ 0.05 & 1.75 $\pm$ 0.06 & 1.90 $\pm$ 0.06 & 2.05 $\pm$ 0.06\\ 
 2.5 & 1.34 $\pm$ 0.05 & 1.34 $\pm$ 0.05 & 1.42 $\pm$ 0.05 & 1.53 $\pm$ 0.05 & 1.67 $\pm$ 0.05 & 1.81 $\pm$ 0.06 & 1.95 $\pm$ 0.06 & 2.09 $\pm$ 0.07\\ 
 3.0 & 1.51 $\pm$ 0.05 & 1.50 $\pm$ 0.05 & 1.54 $\pm$ 0.05 & 1.64 $\pm$ 0.05 & 1.75 $\pm$ 0.06 & 1.88 $\pm$ 0.06 & 2.01 $\pm$ 0.06 & 2.14 $\pm$ 0.07\\ 
 3.5 & 1.67 $\pm$ 0.06 & 1.65 $\pm$ 0.05 & 1.68 $\pm$ 0.05 & 1.75 $\pm$ 0.06 & 1.85 $\pm$ 0.06 & 1.96 $\pm$ 0.06 & 2.08 $\pm$ 0.06 & 2.20 $\pm$ 0.07\\ 
 4.0 & 1.81 $\pm$ 0.06 & 1.79 $\pm$ 0.06 & 1.82 $\pm$ 0.06 & 1.87 $\pm$ 0.06 & 1.96 $\pm$ 0.06 & 2.05 $\pm$ 0.06 & 2.16 $\pm$ 0.07 & 2.26 $\pm$ 0.07\\ 
 4.5 & 1.95 $\pm$ 0.06 & 1.94 $\pm$ 0.06 & 1.96 $\pm$ 0.06 & 2.00 $\pm$ 0.06 & 2.07 $\pm$ 0.06 & 2.15 $\pm$ 0.07 & 2.24 $\pm$ 0.07 & 2.33 $\pm$ 0.07\\
\hline
\end{longtable}

\begin{longtable}{c|cccccccc}
\caption{The contaminating fluxes (host galaxy + nearby companions) of 1ES 0806+524 in mJy.}\\
\hline\hline
Aperture & \multicolumn{8}{c}{FWHM}\\
radius   & \multicolumn{8}{c}{(arcsec)}\\
(arcsec) &\\
      & 1.0    & 2.0         & 3.0          &4.0          &5.0          &6.0          &7.0          &8.0\\
\hline
\endfirsthead
 1.0 & 0.26 $\pm$ 0.04 & 0.31 $\pm$ 0.04 & 0.36 $\pm$ 0.03 & 0.41 $\pm$ 0.03 & 0.45 $\pm$ 0.03 & 0.49 $\pm$ 0.03 & 0.52 $\pm$ 0.03 & 0.55 $\pm$ 0.03\\ 
 1.5 & 0.32 $\pm$ 0.04 & 0.35 $\pm$ 0.04 & 0.38 $\pm$ 0.04 & 0.43 $\pm$ 0.03 & 0.46 $\pm$ 0.03 & 0.50 $\pm$ 0.03 & 0.53 $\pm$ 0.03 & 0.56 $\pm$ 0.04\\ 
 2.0 & 0.38 $\pm$ 0.04 & 0.39 $\pm$ 0.04 & 0.41 $\pm$ 0.04 & 0.45 $\pm$ 0.03 & 0.48 $\pm$ 0.03 & 0.51 $\pm$ 0.03 & 0.54 $\pm$ 0.03 & 0.56 $\pm$ 0.04\\ 
 2.5 & 0.43 $\pm$ 0.04 & 0.43 $\pm$ 0.04 & 0.45 $\pm$ 0.04 & 0.47 $\pm$ 0.03 & 0.50 $\pm$ 0.03 & 0.52 $\pm$ 0.03 & 0.55 $\pm$ 0.03 & 0.57 $\pm$ 0.03\\ 
 3.0 & 0.48 $\pm$ 0.04 & 0.47 $\pm$ 0.04 & 0.48 $\pm$ 0.04 & 0.49 $\pm$ 0.03 & 0.51 $\pm$ 0.03 & 0.54 $\pm$ 0.03 & 0.56 $\pm$ 0.03 & 0.58 $\pm$ 0.03\\ 
 3.5 & 0.52 $\pm$ 0.04 & 0.51 $\pm$ 0.04 & 0.51 $\pm$ 0.04 & 0.52 $\pm$ 0.04 & 0.53 $\pm$ 0.04 & 0.55 $\pm$ 0.03 & 0.57 $\pm$ 0.04 & 0.59 $\pm$ 0.03\\ 
 4.0 & 0.55 $\pm$ 0.04 & 0.54 $\pm$ 0.04 & 0.54 $\pm$ 0.04 & 0.54 $\pm$ 0.04 & 0.55 $\pm$ 0.04 & 0.57 $\pm$ 0.03 & 0.59 $\pm$ 0.04 & 0.60 $\pm$ 0.04\\ 
 4.5 & 0.58 $\pm$ 0.04 & 0.57 $\pm$ 0.04 & 0.56 $\pm$ 0.04 & 0.57 $\pm$ 0.04 & 0.57 $\pm$ 0.04 & 0.59 $\pm$ 0.04 & 0.60 $\pm$ 0.04 & 0.61 $\pm$ 0.04\\
\hline
\end{longtable}

\clearpage

\begin{longtable}{c|cccccccc}
\caption{The contaminating fluxes (host galaxy + nearby companions) of 1ES 1011+496 in mJy.}\\
\hline\hline
Aperture & \multicolumn{8}{c}{FWHM}\\
radius   & \multicolumn{8}{c}{(arcsec)}\\
(arcsec) &\\
      & 1.0    & 2.0         & 3.0          &4.0          &5.0          &6.0          &7.0          &8.0\\
\hline
\endfirsthead
 1.0 & 0.26 $\pm$ 0.01 & 0.29 $\pm$ 0.01 & 0.32 $\pm$ 0.01 & 0.34 $\pm$ 0.02 & 0.37 $\pm$ 0.02 & 0.38 $\pm$ 0.02 & 0.40 $\pm$ 0.02 & 0.42 $\pm$ 0.02\\ 
 1.5 & 0.30 $\pm$ 0.01 & 0.31 $\pm$ 0.01 & 0.33 $\pm$ 0.02 & 0.35 $\pm$ 0.02 & 0.37 $\pm$ 0.02 & 0.39 $\pm$ 0.02 & 0.41 $\pm$ 0.02 & 0.42 $\pm$ 0.02\\ 
 2.0 & 0.33 $\pm$ 0.02 & 0.33 $\pm$ 0.02 & 0.35 $\pm$ 0.02 & 0.36 $\pm$ 0.02 & 0.38 $\pm$ 0.02 & 0.40 $\pm$ 0.02 & 0.41 $\pm$ 0.02 & 0.42 $\pm$ 0.02\\ 
 2.5 & 0.36 $\pm$ 0.02 & 0.36 $\pm$ 0.02 & 0.36 $\pm$ 0.02 & 0.38 $\pm$ 0.02 & 0.39 $\pm$ 0.02 & 0.40 $\pm$ 0.02 & 0.42 $\pm$ 0.02 & 0.43 $\pm$ 0.02\\ 
 3.0 & 0.38 $\pm$ 0.02 & 0.38 $\pm$ 0.02 & 0.38 $\pm$ 0.02 & 0.39 $\pm$ 0.02 & 0.40 $\pm$ 0.02 & 0.41 $\pm$ 0.02 & 0.42 $\pm$ 0.02 & 0.43 $\pm$ 0.02\\ 
 3.5 & 0.40 $\pm$ 0.02 & 0.40 $\pm$ 0.02 & 0.40 $\pm$ 0.02 & 0.40 $\pm$ 0.02 & 0.41 $\pm$ 0.02 & 0.42 $\pm$ 0.02 & 0.43 $\pm$ 0.02 & 0.44 $\pm$ 0.02\\ 
 4.0 & 0.42 $\pm$ 0.02 & 0.41 $\pm$ 0.02 & 0.41 $\pm$ 0.02 & 0.41 $\pm$ 0.02 & 0.42 $\pm$ 0.02 & 0.43 $\pm$ 0.02 & 0.43 $\pm$ 0.02 & 0.44 $\pm$ 0.02\\ 
 4.5 & 0.43 $\pm$ 0.02 & 0.43 $\pm$ 0.02 & 0.42 $\pm$ 0.02 & 0.43 $\pm$ 0.02 & 0.43 $\pm$ 0.02 & 0.43 $\pm$ 0.02 & 0.44 $\pm$ 0.02 & 0.45 $\pm$ 0.02\\
\hline
\end{longtable}

\begin{longtable}{c|cccccccc}
\caption{The contaminating fluxes (host galaxy + nearby companions) of 1ES 1028+511 in mJy.}\\
\hline\hline
Aperture & \multicolumn{8}{c}{FWHM}\\
radius   & \multicolumn{8}{c}{(arcsec)}\\
(arcsec) &\\
      & 1.0    & 2.0         & 3.0          &4.0          &5.0          &6.0          &7.0          &8.0\\
\hline
\endfirsthead
 1.0 & 0.04 $\pm$ 0.02 & 0.05 $\pm$ 0.02 & 0.06 $\pm$ 0.02 & 0.06 $\pm$ 0.02 & 0.07 $\pm$ 0.02 & 0.07 $\pm$ 0.02 & 0.08 $\pm$ 0.02 & 0.08 $\pm$ 0.02\\ 
 1.5 & 0.05 $\pm$ 0.02 & 0.05 $\pm$ 0.02 & 0.06 $\pm$ 0.02 & 0.07 $\pm$ 0.02 & 0.07 $\pm$ 0.02 & 0.07 $\pm$ 0.02 & 0.08 $\pm$ 0.02 & 0.08 $\pm$ 0.02\\ 
 2.0 & 0.06 $\pm$ 0.02 & 0.06 $\pm$ 0.02 & 0.06 $\pm$ 0.02 & 0.07 $\pm$ 0.02 & 0.07 $\pm$ 0.02 & 0.08 $\pm$ 0.02 & 0.08 $\pm$ 0.02 & 0.08 $\pm$ 0.02\\ 
 2.5 & 0.07 $\pm$ 0.02 & 0.07 $\pm$ 0.02 & 0.07 $\pm$ 0.02 & 0.07 $\pm$ 0.02 & 0.07 $\pm$ 0.02 & 0.08 $\pm$ 0.02 & 0.08 $\pm$ 0.02 & 0.08 $\pm$ 0.02\\ 
 3.0 & 0.07 $\pm$ 0.02 & 0.07 $\pm$ 0.02 & 0.07 $\pm$ 0.02 & 0.07 $\pm$ 0.02 & 0.08 $\pm$ 0.02 & 0.08 $\pm$ 0.02 & 0.08 $\pm$ 0.02 & 0.08 $\pm$ 0.02\\ 
 3.5 & 0.08 $\pm$ 0.02 & 0.08 $\pm$ 0.02 & 0.08 $\pm$ 0.02 & 0.08 $\pm$ 0.02 & 0.08 $\pm$ 0.02 & 0.08 $\pm$ 0.02 & 0.08 $\pm$ 0.02 & 0.08 $\pm$ 0.02\\ 
 4.0 & 0.08 $\pm$ 0.02 & 0.08 $\pm$ 0.02 & 0.08 $\pm$ 0.02 & 0.08 $\pm$ 0.02 & 0.08 $\pm$ 0.02 & 0.08 $\pm$ 0.02 & 0.08 $\pm$ 0.02 & 0.09 $\pm$ 0.02\\ 
 4.5 & 0.09 $\pm$ 0.02 & 0.08 $\pm$ 0.02 & 0.08 $\pm$ 0.02 & 0.08 $\pm$ 0.02 & 0.08 $\pm$ 0.02 & 0.08 $\pm$ 0.02 & 0.09 $\pm$ 0.02 & 0.09 $\pm$ 0.02\\
\hline
\end{longtable}

\clearpage

\begin{longtable}{c|cccccccc}
\caption{The contaminating fluxes (host galaxy + nearby companions) of Mrk 421 in mJy.}\\
\hline\hline
Aperture & \multicolumn{8}{c}{FWHM}\\
radius   & \multicolumn{8}{c}{(arcsec)}\\
(arcsec) &\\
      & 1.0    & 2.0         & 3.0          &4.0          &5.0          &6.0          &7.0          &8.0\\
\hline
\endfirsthead
 1.0 & 1.3 $\pm$ 0.2 & 1.8 $\pm$ 0.2 & 2.5 $\pm$ 0.3 & 3.2 $\pm$ 0.3 & 3.8 $\pm$ 0.3 & 4.4 $\pm$ 0.3 & 5.0 $\pm$ 0.4 & 5.5 $\pm$ 0.4\\  
 1.5 & 1.9 $\pm$ 0.3 & 2.2 $\pm$ 0.3 & 2.8 $\pm$ 0.3 & 3.4 $\pm$ 0.3 & 4.0 $\pm$ 0.3 & 4.5 $\pm$ 0.3 & 5.1 $\pm$ 0.4 & 5.6 $\pm$ 0.4\\  
 2.0 & 2.6 $\pm$ 0.3 & 2.7 $\pm$ 0.3 & 3.1 $\pm$ 0.3 & 3.7 $\pm$ 0.3 & 4.2 $\pm$ 0.3 & 4.7 $\pm$ 0.4 & 5.2 $\pm$ 0.4 & 5.7 $\pm$ 0.4\\  
 2.5 & 3.2 $\pm$ 0.4 & 3.3 $\pm$ 0.3 & 3.6 $\pm$ 0.3 & 4.0 $\pm$ 0.3 & 4.5 $\pm$ 0.3 & 4.9 $\pm$ 0.4 & 5.4 $\pm$ 0.4 & 5.9 $\pm$ 0.4\\  
 3.0 & 3.9 $\pm$ 0.4 & 3.9 $\pm$ 0.4 & 4.0 $\pm$ 0.4 & 4.4 $\pm$ 0.4 & 4.8 $\pm$ 0.4 & 5.2 $\pm$ 0.4 & 5.6 $\pm$ 0.4 & 6.0 $\pm$ 0.4\\  
 3.5 & 4.5 $\pm$ 0.4 & 4.4 $\pm$ 0.4 & 4.5 $\pm$ 0.4 & 4.8 $\pm$ 0.4 & 5.1 $\pm$ 0.4 & 5.5 $\pm$ 0.4 & 5.8 $\pm$ 0.4 & 6.2 $\pm$ 0.4\\  
 4.0 & 5.1 $\pm$ 0.4 & 5.0 $\pm$ 0.4 & 5.0 $\pm$ 0.4 & 5.2 $\pm$ 0.4 & 5.4 $\pm$ 0.4 & 5.7 $\pm$ 0.4 & 6.1 $\pm$ 0.4 & 6.4 $\pm$ 0.4\\  
 4.5 & 5.7 $\pm$ 0.4 & 5.5 $\pm$ 0.4 & 5.5 $\pm$ 0.4 & 5.6 $\pm$ 0.4 & 5.8 $\pm$ 0.4 & 6.1 $\pm$ 0.4 & 6.3 $\pm$ 0.4 & 6.7 $\pm$ 0.4\\
\hline
\end{longtable}

\begin{longtable}{c|cccccccc}
\caption{The contaminating fluxes (host galaxy + nearby companions) of RGB J1117+202 in mJy.}\\
\hline\hline
Aperture & \multicolumn{8}{c}{FWHM}\\
radius   & \multicolumn{8}{c}{(arcsec)}\\
(arcsec) &\\
      & 1.0    & 2.0         & 3.0          &4.0          &5.0          &6.0          &7.0          &8.0\\
\hline
\endfirsthead
 1.0 & 0.16 $\pm$ 0.03 & 0.20 $\pm$ 0.03 & 0.26 $\pm$ 0.03 & 0.31 $\pm$ 0.03 & 0.36 $\pm$ 0.03 & 0.40 $\pm$ 0.03 & 0.44 $\pm$ 0.03 & 0.48 $\pm$ 0.03\\ 
 1.5 & 0.21 $\pm$ 0.03 & 0.24 $\pm$ 0.03 & 0.28 $\pm$ 0.03 & 0.32 $\pm$ 0.03 & 0.37 $\pm$ 0.03 & 0.41 $\pm$ 0.03 & 0.45 $\pm$ 0.03 & 0.49 $\pm$ 0.03\\ 
 2.0 & 0.27 $\pm$ 0.03 & 0.28 $\pm$ 0.03 & 0.31 $\pm$ 0.03 & 0.35 $\pm$ 0.03 & 0.38 $\pm$ 0.03 & 0.42 $\pm$ 0.03 & 0.46 $\pm$ 0.03 & 0.50 $\pm$ 0.03\\ 
 2.5 & 0.32 $\pm$ 0.03 & 0.32 $\pm$ 0.03 & 0.34 $\pm$ 0.03 & 0.37 $\pm$ 0.03 & 0.40 $\pm$ 0.03 & 0.44 $\pm$ 0.03 & 0.47 $\pm$ 0.03 & 0.51 $\pm$ 0.03\\ 
 3.0 & 0.37 $\pm$ 0.03 & 0.36 $\pm$ 0.03 & 0.37 $\pm$ 0.03 & 0.40 $\pm$ 0.03 & 0.42 $\pm$ 0.03 & 0.46 $\pm$ 0.03 & 0.49 $\pm$ 0.03 & 0.52 $\pm$ 0.03\\ 
 3.5 & 0.41 $\pm$ 0.03 & 0.40 $\pm$ 0.03 & 0.41 $\pm$ 0.03 & 0.42 $\pm$ 0.03 & 0.45 $\pm$ 0.03 & 0.47 $\pm$ 0.03 & 0.51 $\pm$ 0.03 & 0.54 $\pm$ 0.03\\ 
 4.0 & 0.45 $\pm$ 0.03 & 0.44 $\pm$ 0.03 & 0.44 $\pm$ 0.03 & 0.45 $\pm$ 0.03 & 0.47 $\pm$ 0.03 & 0.50 $\pm$ 0.03 & 0.52 $\pm$ 0.03 & 0.55 $\pm$ 0.03\\ 
 4.5 & 0.49 $\pm$ 0.03 & 0.48 $\pm$ 0.03 & 0.47 $\pm$ 0.03 & 0.48 $\pm$ 0.03 & 0.50 $\pm$ 0.03 & 0.52 $\pm$ 0.03 & 0.54 $\pm$ 0.03 & 0.57 $\pm$ 0.03\\
\hline
\end{longtable}

\clearpage

\begin{longtable}{c|ccccccc}
\caption{The contaminating fluxes (host galaxy + nearby companions) of Mrk 180 in mJy.}\\
\hline\hline
Aperture & \multicolumn{7}{c}{FWHM}\\
radius   & \multicolumn{7}{c}{(arcsec)}\\
(arcsec) &\\
      & 2.0         & 3.0          &4.0          &5.0          &6.0          &7.0          &8.0\\
\hline
\endfirsthead
 1.5 & 1.29 $\pm$ 0.06 & 1.53 $\pm$ 0.07 & 1.82 $\pm$ 0.08 & 2.14 $\pm$ 0.09 &  2.5 $\pm$ 0.1  &  2.8 $\pm$ 0.2  &  3.1 $\pm$ 0.2 \\ 
 2.0 & 1.52 $\pm$ 0.07 & 1.70 $\pm$ 0.07 & 1.96 $\pm$ 0.08 & 2.26 $\pm$ 0.09 &  2.6 $\pm$ 0.2  &  2.9 $\pm$ 0.2  &  3.2 $\pm$ 0.2 \\ 
 2.5 & 1.75 $\pm$ 0.07 & 1.89 $\pm$ 0.08 & 2.12 $\pm$ 0.09 &  2.4 $\pm$ 0.1  &  2.7 $\pm$ 0.2  &  3.0 $\pm$ 0.2  &  3.3 $\pm$ 0.2 \\ 
 3.0 & 1.98 $\pm$ 0.08 & 2.09 $\pm$ 0.09 &  2.3 $\pm$ 0.1  &  2.6 $\pm$ 0.2  &  2.8 $\pm$ 0.2  &  3.1 $\pm$ 0.2  &  3.4 $\pm$ 0.2 \\ 
 3.5 & 2.22 $\pm$ 0.09 &  2.3 $\pm$ 0.1  &  2.5 $\pm$ 0.1  &  2.7 $\pm$ 0.2  &  3.0 $\pm$ 0.2  &  3.3 $\pm$ 0.2  &  3.6 $\pm$ 0.2 \\ 
 4.0 &  2.4 $\pm$ 0.1  &  2.6 $\pm$ 0.2  &  2.7 $\pm$ 0.2  &  3.0 $\pm$ 0.2  &  3.2 $\pm$ 0.2  &  3.5 $\pm$ 0.2  &  3.7 $\pm$ 0.2 \\ 
 4.5 &  2.7 $\pm$ 0.2  &  2.8 $\pm$ 0.2  &  3.0 $\pm$ 0.2  &  3.2 $\pm$ 0.2  &  3.4 $\pm$ 0.2  &  3.7 $\pm$ 0.2  &  3.9 $\pm$ 0.2 \\ 
 5.0 &  3.0 $\pm$ 0.2  &  3.2 $\pm$ 0.2  &  3.3 $\pm$ 0.2  &  3.5 $\pm$ 0.2  &  3.7 $\pm$ 0.2  &  3.9 $\pm$ 0.2  &  4.1 $\pm$ 0.2 \\
\hline
\end{longtable}

\begin{longtable}{c|cccccccc}
\caption{The contaminating fluxes (host galaxy + nearby companions) of RGB J1136+676 in mJy.}\\
\hline\hline
Aperture & \multicolumn{8}{c}{FWHM}\\
radius   & \multicolumn{8}{c}{(arcsec)}\\
(arcsec) &\\
      & 1.0    & 2.0         & 3.0          &4.0          &5.0          &6.0          &7.0          &8.0\\
\hline
\endfirsthead
 1.0 & 0.30 $\pm$ 0.02 & 0.36 $\pm$ 0.02 & 0.43 $\pm$ 0.02 & 0.49 $\pm$ 0.02 & 0.54 $\pm$ 0.03 & 0.59 $\pm$ 0.03 & 0.63 $\pm$ 0.03 & 0.67 $\pm$ 0.03\\ 
 1.5 & 0.38 $\pm$ 0.02 & 0.41 $\pm$ 0.02 & 0.46 $\pm$ 0.02 & 0.51 $\pm$ 0.03 & 0.56 $\pm$ 0.03 & 0.60 $\pm$ 0.03 & 0.64 $\pm$ 0.03 & 0.67 $\pm$ 0.03\\ 
 2.0 & 0.46 $\pm$ 0.02 & 0.46 $\pm$ 0.02 & 0.50 $\pm$ 0.03 & 0.54 $\pm$ 0.03 & 0.58 $\pm$ 0.03 & 0.62 $\pm$ 0.03 & 0.65 $\pm$ 0.03 & 0.69 $\pm$ 0.03\\ 
 2.5 & 0.52 $\pm$ 0.03 & 0.52 $\pm$ 0.03 & 0.53 $\pm$ 0.03 & 0.56 $\pm$ 0.03 & 0.60 $\pm$ 0.03 & 0.63 $\pm$ 0.03 & 0.66 $\pm$ 0.03 & 0.70 $\pm$ 0.03\\ 
 3.0 & 0.57 $\pm$ 0.03 & 0.56 $\pm$ 0.03 & 0.57 $\pm$ 0.03 & 0.59 $\pm$ 0.03 & 0.62 $\pm$ 0.03 & 0.65 $\pm$ 0.03 & 0.68 $\pm$ 0.03 & 0.71 $\pm$ 0.03\\ 
 3.5 & 0.62 $\pm$ 0.03 & 0.61 $\pm$ 0.03 & 0.61 $\pm$ 0.03 & 0.62 $\pm$ 0.03 & 0.64 $\pm$ 0.03 & 0.67 $\pm$ 0.03 & 0.69 $\pm$ 0.03 & 0.72 $\pm$ 0.03\\ 
 4.0 & 0.66 $\pm$ 0.03 & 0.65 $\pm$ 0.03 & 0.65 $\pm$ 0.03 & 0.65 $\pm$ 0.03 & 0.67 $\pm$ 0.03 & 0.69 $\pm$ 0.03 & 0.71 $\pm$ 0.03 & 0.74 $\pm$ 0.03\\ 
 4.5 & 0.70 $\pm$ 0.03 & 0.69 $\pm$ 0.03 & 0.68 $\pm$ 0.03 & 0.69 $\pm$ 0.03 & 0.70 $\pm$ 0.03 & 0.71 $\pm$ 0.03 & 0.73 $\pm$ 0.03 & 0.75 $\pm$ 0.04\\
\hline
\end{longtable}

\clearpage

\begin{longtable}{c|ccccccc}
\caption{The contaminating fluxes (host galaxy + nearby companions) of ON 325 in mJy.}\\
\hline\hline
Aperture & \multicolumn{7}{c}{FWHM}\\
radius   & \multicolumn{7}{c}{(arcsec)}\\
(arcsec) &\\
      & 2.0         & 3.0          &4.0          &5.0          &6.0          &7.0          &8.0\\
\hline
\endfirsthead
 1.5 & 0.54 $\pm$ 0.09 & 0.59 $\pm$ 0.09 & 0.64 $\pm$ 0.09 & 0.69 $\pm$ 0.09 & 0.74 $\pm$ 0.09 & 0.78 $\pm$ 0.09 & 0.81 $\pm$ 0.09\\ 
 2.0 & 0.59 $\pm$ 0.09 & 0.63 $\pm$ 0.09 & 0.67 $\pm$ 0.09 & 0.71 $\pm$ 0.09 & 0.75 $\pm$ 0.09 & 0.79 $\pm$ 0.09 & 0.82 $\pm$ 0.09\\ 
 2.5 & 0.65 $\pm$ 0.09 & 0.67 $\pm$ 0.09 & 0.70 $\pm$ 0.09 & 0.74 $\pm$ 0.09 & 0.77 $\pm$ 0.09 & 0.80 $\pm$ 0.09 & 0.83 $\pm$ 0.09\\ 
 3.0 & 0.70 $\pm$ 0.09 & 0.71 $\pm$ 0.09 & 0.73 $\pm$ 0.09 & 0.76 $\pm$ 0.09 & 0.79 $\pm$ 0.09 & 0.82 $\pm$ 0.09 & 0.85 $\pm$ 0.09\\ 
 3.5 & 0.75 $\pm$ 0.09 & 0.75 $\pm$ 0.09 & 0.77 $\pm$ 0.09 & 0.79 $\pm$ 0.09 & 0.81 $\pm$ 0.09 & 0.84 $\pm$ 0.09 & 0.86 $\pm$ 0.09\\ 
 4.0 &  0.8 $\pm$ 0.1  &  0.8 $\pm$ 0.1  & 0.80 $\pm$ 0.09 & 0.81 $\pm$ 0.09 & 0.83 $\pm$ 0.09 & 0.85 $\pm$ 0.09 & 0.87 $\pm$ 0.09\\ 
 4.5 &  0.8 $\pm$ 0.1  &  0.8 $\pm$ 0.1  &  0.8 $\pm$ 0.1  & 0.84 $\pm$ 0.09 & 0.85 $\pm$ 0.09 & 0.87 $\pm$ 0.09 & 0.89 $\pm$ 0.09\\ 
 5.0 &  0.9 $\pm$ 0.1  &  0.9 $\pm$ 0.1  &  0.9 $\pm$ 0.1  & 0.87 $\pm$ 0.09 & 0.88 $\pm$ 0.09 & 0.89 $\pm$ 0.09 & 0.91 $\pm$ 0.09\\
\hline
\end{longtable}

\begin{longtable}{c|ccccccc}
\caption{The contaminating fluxes (host galaxy + nearby companions) of 1ES 1218+304 in mJy.}\\
\hline\hline
Aperture & \multicolumn{7}{c}{FWHM}\\
radius   & \multicolumn{7}{c}{(arcsec)}\\
(arcsec) &\\
      & 2.0         & 3.0          &4.0          &5.0          &6.0          &7.0          &8.0\\
\hline
\endfirsthead
 2.0 & 0.21 $\pm$ 0.02 & 0.22 $\pm$ 0.02 & 0.25 $\pm$ 0.02 & 0.27 $\pm$ 0.02 & 0.29 $\pm$ 0.02 & 0.32 $\pm$ 0.02 & 0.34 $\pm$ 0.02\\ 
 2.5 & 0.23 $\pm$ 0.02 & 0.24 $\pm$ 0.02 & 0.26 $\pm$ 0.02 & 0.28 $\pm$ 0.02 & 0.30 $\pm$ 0.02 & 0.33 $\pm$ 0.02 & 0.35 $\pm$ 0.02\\ 
 3.0 & 0.26 $\pm$ 0.02 & 0.26 $\pm$ 0.02 & 0.28 $\pm$ 0.02 & 0.29 $\pm$ 0.02 & 0.31 $\pm$ 0.02 & 0.33 $\pm$ 0.02 & 0.36 $\pm$ 0.02\\ 
 3.5 & 0.28 $\pm$ 0.02 & 0.28 $\pm$ 0.02 & 0.29 $\pm$ 0.02 & 0.31 $\pm$ 0.02 & 0.32 $\pm$ 0.02 & 0.34 $\pm$ 0.02 & 0.37 $\pm$ 0.02\\ 
 4.0 & 0.30 $\pm$ 0.02 & 0.30 $\pm$ 0.02 & 0.31 $\pm$ 0.02 & 0.32 $\pm$ 0.02 & 0.34 $\pm$ 0.02 & 0.36 $\pm$ 0.02 & 0.38 $\pm$ 0.02\\ 
 4.5 & 0.32 $\pm$ 0.02 & 0.32 $\pm$ 0.02 & 0.32 $\pm$ 0.02 & 0.33 $\pm$ 0.02 & 0.35 $\pm$ 0.02 & 0.37 $\pm$ 0.02 & 0.39 $\pm$ 0.02\\ 
 5.0 & 0.34 $\pm$ 0.02 & 0.34 $\pm$ 0.02 & 0.34 $\pm$ 0.02 & 0.35 $\pm$ 0.02 & 0.36 $\pm$ 0.02 & 0.38 $\pm$ 0.02 & 0.40 $\pm$ 0.02\\ 
 5.5 & 0.36 $\pm$ 0.02 & 0.35 $\pm$ 0.02 & 0.35 $\pm$ 0.02 & 0.36 $\pm$ 0.02 & 0.38 $\pm$ 0.02 & 0.39 $\pm$ 0.02 & 0.42 $\pm$ 0.02\\
\hline
\end{longtable}

\clearpage

\begin{longtable}{c|cccccccc}
\caption{The contaminating fluxes (host galaxy + nearby companions) of RGB J1417+257 in mJy.}\\
\hline\hline
Aperture & \multicolumn{8}{c}{FWHM}\\
radius   & \multicolumn{8}{c}{(arcsec)}\\
(arcsec) &\\
      & 1.0    & 2.0         & 3.0          &4.0          &5.0          &6.0          &7.0          &8.0\\
\hline
\endfirsthead
 1.0 & 0.10 $\pm$ 0.03 & 0.14 $\pm$ 0.04 & 0.18 $\pm$ 0.04 & 0.21 $\pm$ 0.04 & 0.25 $\pm$ 0.04 & 0.29 $\pm$ 0.05 & 0.32 $\pm$ 0.05 & 0.35 $\pm$ 0.05\\ 
 1.5 & 0.14 $\pm$ 0.04 & 0.16 $\pm$ 0.04 & 0.19 $\pm$ 0.04 & 0.23 $\pm$ 0.04 & 0.26 $\pm$ 0.05 & 0.29 $\pm$ 0.05 & 0.33 $\pm$ 0.05 & 0.36 $\pm$ 0.05\\ 
 2.0 & 0.18 $\pm$ 0.04 & 0.19 $\pm$ 0.04 & 0.21 $\pm$ 0.04 & 0.24 $\pm$ 0.04 & 0.27 $\pm$ 0.05 & 0.31 $\pm$ 0.05 & 0.34 $\pm$ 0.05 & 0.36 $\pm$ 0.05\\ 
 2.5 & 0.22 $\pm$ 0.04 & 0.22 $\pm$ 0.04 & 0.24 $\pm$ 0.04 & 0.26 $\pm$ 0.05 & 0.29 $\pm$ 0.05 & 0.32 $\pm$ 0.05 & 0.35 $\pm$ 0.05 & 0.37 $\pm$ 0.05\\ 
 3.0 & 0.25 $\pm$ 0.05 & 0.25 $\pm$ 0.04 & 0.26 $\pm$ 0.05 & 0.28 $\pm$ 0.05 & 0.31 $\pm$ 0.05 & 0.33 $\pm$ 0.05 & 0.36 $\pm$ 0.05 & 0.38 $\pm$ 0.05\\ 
 3.5 & 0.29 $\pm$ 0.05 & 0.28 $\pm$ 0.05 & 0.29 $\pm$ 0.05 & 0.31 $\pm$ 0.05 & 0.33 $\pm$ 0.05 & 0.35 $\pm$ 0.05 & 0.37 $\pm$ 0.05 & 0.39 $\pm$ 0.05\\ 
 4.0 & 0.32 $\pm$ 0.05 & 0.32 $\pm$ 0.05 & 0.32 $\pm$ 0.05 & 0.33 $\pm$ 0.05 & 0.35 $\pm$ 0.05 & 0.37 $\pm$ 0.05 & 0.39 $\pm$ 0.05 & 0.40 $\pm$ 0.05\\ 
 4.5 & 0.35 $\pm$ 0.05 & 0.35 $\pm$ 0.05 & 0.35 $\pm$ 0.05 & 0.36 $\pm$ 0.05 & 0.37 $\pm$ 0.05 & 0.39 $\pm$ 0.05 & 0.40 $\pm$ 0.05 & 0.42 $\pm$ 0.05\\
\hline
\end{longtable}

\begin{longtable}{c|cccccccc}
\caption{The contaminating fluxes (host galaxy + nearby companions) of 1ES 1426+428 in mJy.}\\
\hline\hline
Aperture & \multicolumn{8}{c}{FWHM}\\
radius   & \multicolumn{8}{c}{(arcsec)}\\
(arcsec) &\\
      & 1.0    & 2.0         & 3.0          &4.0          &5.0          &6.0          &7.0          &8.0\\
\hline
\endfirsthead
 1.0 & 0.29 $\pm$ 0.02 & 0.36 $\pm$ 0.02 & 0.43 $\pm$ 0.02 & 0.50 $\pm$ 0.02 & 0.56 $\pm$ 0.02 & 0.61 $\pm$ 0.02 & 0.65 $\pm$ 0.03 & 0.69 $\pm$ 0.03\\ 
 1.5 & 0.38 $\pm$ 0.02 & 0.41 $\pm$ 0.02 & 0.47 $\pm$ 0.02 & 0.52 $\pm$ 0.02 & 0.57 $\pm$ 0.02 & 0.62 $\pm$ 0.03 & 0.66 $\pm$ 0.03 & 0.70 $\pm$ 0.03\\ 
 2.0 & 0.46 $\pm$ 0.02 & 0.47 $\pm$ 0.02 & 0.51 $\pm$ 0.02 & 0.55 $\pm$ 0.02 & 0.59 $\pm$ 0.02 & 0.64 $\pm$ 0.03 & 0.67 $\pm$ 0.03 & 0.71 $\pm$ 0.03\\ 
 2.5 & 0.53 $\pm$ 0.02 & 0.53 $\pm$ 0.02 & 0.55 $\pm$ 0.02 & 0.58 $\pm$ 0.02 & 0.62 $\pm$ 0.03 & 0.65 $\pm$ 0.03 & 0.69 $\pm$ 0.03 & 0.72 $\pm$ 0.03\\ 
 3.0 & 0.59 $\pm$ 0.02 & 0.58 $\pm$ 0.02 & 0.59 $\pm$ 0.02 & 0.61 $\pm$ 0.02 & 0.64 $\pm$ 0.03 & 0.67 $\pm$ 0.03 & 0.70 $\pm$ 0.03 & 0.73 $\pm$ 0.03\\ 
 3.5 & 0.65 $\pm$ 0.03 & 0.63 $\pm$ 0.03 & 0.63 $\pm$ 0.03 & 0.65 $\pm$ 0.03 & 0.67 $\pm$ 0.03 & 0.69 $\pm$ 0.03 & 0.72 $\pm$ 0.03 & 0.75 $\pm$ 0.03\\ 
 4.0 & 0.69 $\pm$ 0.03 & 0.68 $\pm$ 0.03 & 0.67 $\pm$ 0.03 & 0.68 $\pm$ 0.03 & 0.70 $\pm$ 0.03 & 0.72 $\pm$ 0.03 & 0.74 $\pm$ 0.03 & 0.76 $\pm$ 0.03\\ 
 4.5 & 0.74 $\pm$ 0.03 & 0.72 $\pm$ 0.03 & 0.71 $\pm$ 0.03 & 0.71 $\pm$ 0.03 & 0.72 $\pm$ 0.03 & 0.74 $\pm$ 0.03 & 0.76 $\pm$ 0.03 & 0.78 $\pm$ 0.03\\
\hline
\end{longtable}

\clearpage

\begin{longtable}{c|ccccccc}
\caption{The contaminating fluxes (host galaxy + nearby companions) of 1ES 1544+820 in mJy.}\\
\hline\hline
Aperture & \multicolumn{7}{c}{FWHM}\\
radius   & \multicolumn{7}{c}{(arcsec)}\\
(arcsec) &\\
      & 2.0         & 3.0          &4.0          &5.0          &6.0          &7.0          &8.0\\
\hline
\endfirsthead
 2.0 & 0.01 $\pm$ 0.01 & 0.02 $\pm$ 0.01 & 0.04 $\pm$ 0.01 & 0.06 $\pm$ 0.01 & 0.08 $\pm$ 0.01 & 0.10 $\pm$ 0.01 & 0.12 $\pm$ 0.01\\ 
 2.5 & 0.02 $\pm$ 0.01 & 0.03 $\pm$ 0.01 & 0.05 $\pm$ 0.01 & 0.07 $\pm$ 0.01 & 0.09 $\pm$ 0.01 & 0.11 $\pm$ 0.01 & 0.12 $\pm$ 0.01\\ 
 3.0 & 0.03 $\pm$ 0.01 & 0.05 $\pm$ 0.01 & 0.07 $\pm$ 0.01 & 0.08 $\pm$ 0.01 & 0.10 $\pm$ 0.01 & 0.11 $\pm$ 0.01 & 0.13 $\pm$ 0.01\\ 
 3.5 & 0.06 $\pm$ 0.01 & 0.07 $\pm$ 0.01 & 0.09 $\pm$ 0.01 & 0.10 $\pm$ 0.01 & 0.11 $\pm$ 0.01 & 0.12 $\pm$ 0.01 & 0.13 $\pm$ 0.01\\ 
 4.0 & 0.09 $\pm$ 0.01 & 0.10 $\pm$ 0.01 & 0.10 $\pm$ 0.01 & 0.11 $\pm$ 0.01 & 0.12 $\pm$ 0.01 & 0.13 $\pm$ 0.01 & 0.14 $\pm$ 0.01\\ 
 4.5 & 0.13 $\pm$ 0.01 & 0.12 $\pm$ 0.01 & 0.12 $\pm$ 0.01 & 0.13 $\pm$ 0.01 & 0.13 $\pm$ 0.01 & 0.14 $\pm$ 0.01 & 0.15 $\pm$ 0.01\\ 
 5.0 & 0.17 $\pm$ 0.01 & 0.15 $\pm$ 0.01 & 0.14 $\pm$ 0.01 & 0.14 $\pm$ 0.01 & 0.15 $\pm$ 0.01 & 0.15 $\pm$ 0.01 & 0.15 $\pm$ 0.01\\ 
 5.5 & 0.19 $\pm$ 0.01 & 0.17 $\pm$ 0.01 & 0.16 $\pm$ 0.01 & 0.16 $\pm$ 0.01 & 0.16 $\pm$ 0.01 & 0.16 $\pm$ 0.01 & 0.16 $\pm$ 0.01\\
\hline
\end{longtable}

\begin{longtable}{c|cccccccc}
\caption{The contaminating fluxes (host galaxy + nearby companions) of Mrk 501 in mJy.}\\
\hline\hline
Aperture & \multicolumn{8}{c}{FWHM}\\
radius   & \multicolumn{8}{c}{(arcsec)}\\
(arcsec) &\\
      & 1.0    & 2.0         & 3.0          &4.0          &5.0          &6.0          &7.0          &8.0\\
\hline
\endfirsthead
 1.0 &  3.4 $\pm$ 0.2 & 4.1 $\pm$ 0.2 & 5.1 $\pm$ 0.2 & 6.0 $\pm$ 0.2 & 6.8 $\pm$ 0.2 & 7.5 $\pm$ 0.2 & 8.2 $\pm$ 0.2 & 8.9 $\pm$ 0.2\\ 
 1.5 &  4.4 $\pm$ 0.2 & 4.8 $\pm$ 0.2 & 5.5 $\pm$ 0.2 & 6.3 $\pm$ 0.2 & 7.0 $\pm$ 0.2 & 7.7 $\pm$ 0.2 & 8.4 $\pm$ 0.2 & 9.0 $\pm$ 0.2\\ 
 2.0 &  5.3 $\pm$ 0.2 & 5.5 $\pm$ 0.2 & 6.0 $\pm$ 0.2 & 6.6 $\pm$ 0.2 & 7.3 $\pm$ 0.2 & 7.9 $\pm$ 0.2 & 8.6 $\pm$ 0.2 & 9.2 $\pm$ 0.2\\ 
 2.5 &  6.2 $\pm$ 0.2 & 6.2 $\pm$ 0.2 & 6.6 $\pm$ 0.2 & 7.1 $\pm$ 0.2 & 7.6 $\pm$ 0.2 & 8.2 $\pm$ 0.2 & 8.8 $\pm$ 0.2 & 9.3 $\pm$ 0.2\\ 
 3.0 &  7.1 $\pm$ 0.2 & 7.0 $\pm$ 0.2 & 7.2 $\pm$ 0.2 & 7.5 $\pm$ 0.2 & 8.0 $\pm$ 0.2 & 8.5 $\pm$ 0.2 & 9.0 $\pm$ 0.2 & 9.5 $\pm$ 0.2\\ 
 3.5 &  7.8 $\pm$ 0.2 & 7.7 $\pm$ 0.2 & 7.8 $\pm$ 0.2 & 8.0 $\pm$ 0.2 & 8.4 $\pm$ 0.2 & 8.8 $\pm$ 0.2 & 9.3 $\pm$ 0.2 & 9.8 $\pm$ 0.2\\ 
 4.0 &  8.5 $\pm$ 0.2 & 8.4 $\pm$ 0.2 & 8.4 $\pm$ 0.2 & 8.5 $\pm$ 0.2 & 8.8 $\pm$ 0.2 & 9.2 $\pm$ 0.2 & 9.6 $\pm$ 0.2 &10.0 $\pm$ 0.3\\ 
 4.5 &  9.2 $\pm$ 0.2 & 9.0 $\pm$ 0.2 & 8.9 $\pm$ 0.2 & 9.1 $\pm$ 0.2 & 9.3 $\pm$ 0.2 & 9.6 $\pm$ 0.2 & 9.9 $\pm$ 0.3 &10.3 $\pm$ 0.3\\
\hline
\end{longtable}

\clearpage

\begin{longtable}{c|cccccc}
\caption{The contaminating fluxes (host galaxy + nearby companions) of OT 546 in mJy.}\\
\hline\hline
Aperture & \multicolumn{6}{c}{FWHM}\\
radius   & \multicolumn{6}{c}{(arcsec)}\\
(arcsec) &\\
      & 3.0          &4.0          &5.0          &6.0          &7.0          &8.0\\
\hline
\endfirsthead
 2.5 & 0.70 $\pm$ 0.05 & 0.75 $\pm$ 0.05 & 0.81 $\pm$ 0.05 & 0.87 $\pm$ 0.05 & 0.93 $\pm$ 0.05 & 0.98 $\pm$ 0.05\\ 
 3.0 & 0.77 $\pm$ 0.05 & 0.80 $\pm$ 0.05 & 0.85 $\pm$ 0.05 & 0.90 $\pm$ 0.05 & 0.95 $\pm$ 0.05 & 1.00 $\pm$ 0.05\\ 
 3.5 & 0.83 $\pm$ 0.05 & 0.86 $\pm$ 0.05 & 0.89 $\pm$ 0.05 & 0.93 $\pm$ 0.05 & 0.98 $\pm$ 0.05 & 1.02 $\pm$ 0.05\\ 
 4.0 & 0.90 $\pm$ 0.06 & 0.91 $\pm$ 0.05 & 0.94 $\pm$ 0.05 & 0.97 $\pm$ 0.05 & 1.01 $\pm$ 0.05 & 1.04 $\pm$ 0.05\\ 
 4.5 & 0.96 $\pm$ 0.06 & 0.96 $\pm$ 0.06 & 0.98 $\pm$ 0.05 & 1.01 $\pm$ 0.05 & 1.04 $\pm$ 0.05 & 1.07 $\pm$ 0.05\\ 
 5.0 & 1.01 $\pm$ 0.06 & 1.01 $\pm$ 0.06 & 1.02 $\pm$ 0.06 & 1.04 $\pm$ 0.05 & 1.07 $\pm$ 0.05 & 1.10 $\pm$ 0.05\\ 
 5.5 & 1.07 $\pm$ 0.06 & 1.06 $\pm$ 0.06 & 1.07 $\pm$ 0.06 & 1.08 $\pm$ 0.06 & 1.10 $\pm$ 0.05 & 1.13 $\pm$ 0.05\\ 
 6.0 & 1.12 $\pm$ 0.06 & 1.11 $\pm$ 0.06 & 1.11 $\pm$ 0.06 & 1.12 $\pm$ 0.06 & 1.13 $\pm$ 0.05 & 1.15 $\pm$ 0.06\\
\hline
\end{longtable}

\begin{longtable}{c|cccccccc}
\caption{The contaminating fluxes (host galaxy + nearby companions) of 1ES 1959+650 in mJy.}\\
\hline\hline
Aperture & \multicolumn{8}{c}{FWHM}\\
radius   & \multicolumn{8}{c}{(arcsec)}\\
(arcsec) &\\
      & 1.0    & 2.0         & 3.0          &4.0          &5.0          &6.0          &7.0          &8.0\\
\hline
\endfirsthead
 1.0 & 0.25 $\pm$ 0.01 & 0.37 $\pm$ 0.01 & 0.51 $\pm$ 0.02 & 0.65 $\pm$ 0.02 & 0.79 $\pm$ 0.02 & 0.92 $\pm$ 0.02 & 1.04 $\pm$ 0.03 & 1.16 $\pm$ 0.03\\ 
 1.5 & 0.38 $\pm$ 0.02 & 0.46 $\pm$ 0.02 & 0.57 $\pm$ 0.02 & 0.70 $\pm$ 0.02 & 0.83 $\pm$ 0.02 & 0.95 $\pm$ 0.02 & 1.07 $\pm$ 0.03 & 1.18 $\pm$ 0.03\\ 
 2.0 & 0.53 $\pm$ 0.02 & 0.56 $\pm$ 0.02 & 0.65 $\pm$ 0.02 & 0.76 $\pm$ 0.02 & 0.87 $\pm$ 0.02 & 0.99 $\pm$ 0.03 & 1.10 $\pm$ 0.03 & 1.20 $\pm$ 0.03\\ 
 2.5 & 0.67 $\pm$ 0.02 & 0.68 $\pm$ 0.02 & 0.74 $\pm$ 0.02 & 0.83 $\pm$ 0.02 & 0.93 $\pm$ 0.02 & 1.03 $\pm$ 0.03 & 1.14 $\pm$ 0.03 & 1.24 $\pm$ 0.03\\ 
 3.0 & 0.81 $\pm$ 0.02 & 0.80 $\pm$ 0.02 & 0.84 $\pm$ 0.02 & 0.91 $\pm$ 0.02 & 0.99 $\pm$ 0.03 & 1.08 $\pm$ 0.03 & 1.18 $\pm$ 0.03 & 1.27 $\pm$ 0.03\\ 
 3.5 & 0.94 $\pm$ 0.02 & 0.93 $\pm$ 0.02 & 0.94 $\pm$ 0.02 & 0.99 $\pm$ 0.03 & 1.06 $\pm$ 0.03 & 1.14 $\pm$ 0.03 & 1.23 $\pm$ 0.03 & 1.32 $\pm$ 0.03\\ 
 4.0 & 1.07 $\pm$ 0.03 & 1.04 $\pm$ 0.03 & 1.05 $\pm$ 0.03 & 1.08 $\pm$ 0.03 & 1.14 $\pm$ 0.03 & 1.21 $\pm$ 0.03 & 1.28 $\pm$ 0.03 & 1.36 $\pm$ 0.03\\ 
 4.5 & 1.18 $\pm$ 0.03 & 1.16 $\pm$ 0.03 & 1.15 $\pm$ 0.03 & 1.17 $\pm$ 0.03 & 1.22 $\pm$ 0.03 & 1.27 $\pm$ 0.03 & 1.34 $\pm$ 0.03 & 1.41 $\pm$ 0.03\\
\hline
\end{longtable}

\clearpage

\begin{longtable}{c|cccccccc}
\caption{The contaminating fluxes (host galaxy + nearby companions) of BL Lac in mJy.}\\
\hline\hline
Aperture & \multicolumn{8}{c}{FWHM}\\
radius   & \multicolumn{8}{c}{(arcsec)}\\
(arcsec) &\\
      & 1.0    & 2.0         & 3.0          &4.0          &5.0          &6.0          &7.0          &8.0\\
\hline
\endfirsthead
 1.0 & 0.35 $\pm$ 0.02 & 0.45 $\pm$ 0.02 & 0.56 $\pm$ 0.02 & 0.67 $\pm$ 0.02 & 0.77 $\pm$ 0.03 & 0.86 $\pm$ 0.03 & 0.94 $\pm$ 0.04 & 1.01 $\pm$ 0.06\\ 
 1.5 & 0.47 $\pm$ 0.02 & 0.53 $\pm$ 0.02 & 0.61 $\pm$ 0.02 & 0.71 $\pm$ 0.02 & 0.80 $\pm$ 0.03 & 0.88 $\pm$ 0.03 & 0.96 $\pm$ 0.04 & 1.03 $\pm$ 0.06\\ 
 2.0 & 0.60 $\pm$ 0.02 & 0.62 $\pm$ 0.02 & 0.68 $\pm$ 0.02 & 0.75 $\pm$ 0.03 & 0.83 $\pm$ 0.03 & 0.91 $\pm$ 0.03 & 0.98 $\pm$ 0.04 & 1.05 $\pm$ 0.06\\ 
 2.5 & 0.71 $\pm$ 0.03 & 0.71 $\pm$ 0.03 & 0.74 $\pm$ 0.03 & 0.80 $\pm$ 0.03 & 0.87 $\pm$ 0.03 & 0.94 $\pm$ 0.03 & 1.00 $\pm$ 0.04 & 1.07 $\pm$ 0.06\\ 
 3.0 & 0.81 $\pm$ 0.03 & 0.80 $\pm$ 0.03 & 0.82 $\pm$ 0.03 & 0.86 $\pm$ 0.03 & 0.91 $\pm$ 0.03 & 0.97 $\pm$ 0.03 & 1.03 $\pm$ 0.05 & 1.09 $\pm$ 0.06\\ 
 3.5 & 0.90 $\pm$ 0.03 & 0.88 $\pm$ 0.03 & 0.89 $\pm$ 0.03 & 0.92 $\pm$ 0.03 & 0.96 $\pm$ 0.03 & 1.01 $\pm$ 0.04 & 1.07 $\pm$ 0.05 & 1.12 $\pm$ 0.07\\ 
 4.0 & 0.98 $\pm$ 0.03 & 0.96 $\pm$ 0.03 & 0.96 $\pm$ 0.03 & 0.98 $\pm$ 0.03 & 1.01 $\pm$ 0.03 & 1.06 $\pm$ 0.04 & 1.10 $\pm$ 0.05 & 1.15 $\pm$ 0.07\\ 
 4.5 & 1.06 $\pm$ 0.03 & 1.04 $\pm$ 0.03 & 1.03 $\pm$ 0.03 & 1.04 $\pm$ 0.03 & 1.07 $\pm$ 0.03 & 1.10 $\pm$ 0.04 & 1.14 $\pm$ 0.05 & 1.18 $\pm$ 0.08\\
\hline
\end{longtable}

\begin{longtable}{c|cccccccc}
\caption{The contaminating fluxes (host galaxy + nearby companions) of 1ES 2344+514 in mJy.}\\
\hline\hline
Aperture & \multicolumn{8}{c}{FWHM}\\
radius   & \multicolumn{8}{c}{(arcsec)}\\
(arcsec) &\\
      & 1.0    & 2.0         & 3.0          &4.0          &5.0          &6.0          &7.0          &8.0\\
\hline
\endfirsthead
 1.0 & 0.88 $\pm$ 0.03 & 1.13 $\pm$ 0.03 & 1.44 $\pm$ 0.03 & 1.73 $\pm$ 0.03 & 2.00 $\pm$ 0.03 & 2.25 $\pm$ 0.04 & 2.49 $\pm$ 0.04 & 2.72 $\pm$ 0.04\\ 
 1.5 & 1.20 $\pm$ 0.03 & 1.34 $\pm$ 0.03 & 1.57 $\pm$ 0.03 & 1.83 $\pm$ 0.03 & 2.07 $\pm$ 0.03 & 2.31 $\pm$ 0.04 & 2.54 $\pm$ 0.04 & 2.76 $\pm$ 0.04\\ 
 2.0 & 1.52 $\pm$ 0.03 & 1.57 $\pm$ 0.03 & 1.74 $\pm$ 0.03 & 1.95 $\pm$ 0.03 & 2.17 $\pm$ 0.04 & 2.39 $\pm$ 0.04 & 2.60 $\pm$ 0.04 & 2.82 $\pm$ 0.04\\ 
 2.5 & 1.81 $\pm$ 0.03 & 1.82 $\pm$ 0.03 & 1.93 $\pm$ 0.03 & 2.09 $\pm$ 0.04 & 2.28 $\pm$ 0.04 & 2.48 $\pm$ 0.04 & 2.68 $\pm$ 0.04 & 2.88 $\pm$ 0.04\\ 
 3.0 & 2.09 $\pm$ 0.04 & 2.06 $\pm$ 0.04 & 2.12 $\pm$ 0.04 & 2.24 $\pm$ 0.04 & 2.40 $\pm$ 0.04 & 2.58 $\pm$ 0.04 & 2.76 $\pm$ 0.04 & 2.95 $\pm$ 0.04\\ 
 3.5 & 2.34 $\pm$ 0.04 & 2.29 $\pm$ 0.04 & 2.32 $\pm$ 0.04 & 2.41 $\pm$ 0.04 & 2.54 $\pm$ 0.04 & 2.69 $\pm$ 0.04 & 2.86 $\pm$ 0.04 & 3.03 $\pm$ 0.04\\ 
 4.0 & 2.57 $\pm$ 0.04 & 2.52 $\pm$ 0.04 & 2.52 $\pm$ 0.04 & 2.58 $\pm$ 0.04 & 2.68 $\pm$ 0.04 & 2.81 $\pm$ 0.04 & 2.96 $\pm$ 0.04 & 3.12 $\pm$ 0.04\\ 
 4.5 & 2.78 $\pm$ 0.04 & 2.73 $\pm$ 0.04 & 2.71 $\pm$ 0.04 & 2.75 $\pm$ 0.04 & 2.83 $\pm$ 0.04 & 2.94 $\pm$ 0.04 & 3.07 $\pm$ 0.04 & 3.22 $\pm$ 0.04\\
\hline
\end{longtable}

\end{appendix}

\end{document}